\documentclass[english,usenatbib,useAMS]{mn2e}
\usepackage[T1]{fontenc}
\usepackage[latin9]{inputenc}
\usepackage{amssymb}
\usepackage{graphicx}
\usepackage{esint}

\makeatletter

\pdfpageheight\paperheight
\pdfpagewidth\paperwidth

\providecommand{\tabularnewline}{\\}

\usepackage{bm}\usepackage{fixltx2e}\usepackage{astrobib_mnras2e}
\usepackage[a4paper]{geometry}
\usepackage{hyperref}

\title[BOSS DR9 LOWZ galaxy clustering]{The clustering of galaxies in the SDSS-III Baryon Oscillation Spectroscopic Survey: the low redshift sample}

\author[Parejko et al.]{\parbox{\textwidth}{
John K. Parejko$^1$\thanks{E-mail: john.parejko@yale.edu},
Tomomi Sunayama$^1$,
Nikhil Padmanabhan$^1$,
David A. Wake$^1$,
Andreas A.  Berlind$^2$,
Dmitry Bizyaev$^3$,
Michael Blanton$^4$,
Adam S. Bolton$^5$,
Frank van den Bosch$^1$,
Jon Brinkmann$^3$,
Joel R. Brownstein$^5$,
Luiz Alberto Nicolaci da Costa$^6$,
Daniel J. Eisenstein$^7$,
Hong Guo$^9$,
Eyal Kazin$^{9}$,
Marcio Maia$^6$,
Elena Malanushenko$^3$,
Claudia Maraston$^{10}$,
Cameron K. McBride$^{2,8}$,
Robert C. Nichol$^{10}$,
Daniel J. Oravetz$^3$,
Kaike Pan$^3$,
Will J. Percival$^{10}$,
Francisco Prada$^{11,12,13}$,
Ashley J. Ross$^{10}$,
Nicholas P. Ross$^{14}$,
David J. Schlegel$^{15}$,
Don Schneider$^{16,17}$,
Audrey E. Simmons$^3$,
Ramin Skibba$^{18}$,
Jeremy Tinker$^4$,
Rita Tojeiro$^{10}$,
Benjamin A. Weaver$^{4}$,
Andrew Wetzel$^1$,
Martin White$^{19,15}$,
David H. Weinberg$^{20}$,
Daniel Thomas$^{10}$,
Idit Zehavi$^8$,
Zheng Zheng$^5$
} \vspace*{4pt} \\
\\$^1$Department of Physics, Yale University, 260 Whitney Ave, New Haven, CT 06520, USA
\\$^2$Department of Physics and Astronomy, Vanderbilt University, Nashville, TN, USA
\\$^3$Apache Point Observatory, P.O. Box 59, Sunspot, NM 88349-0059, USA
\\$^4$Center for Cosmology and Particle Physics, New York University, NY, 10003, USA
\\$^5$Department of Physics and Astronomy, The University of Utah, Salt Lake City, UT 84112, USA
\\$^7$Observatorio Nacional, R. Gal. Jose Cristino 77, S\"{a}o Crist\'{o}v\"{a}o, BR Rio de Janeiro, RJ 20921-400, Brazil
\\$^8$Harvard-Smithsonian Center for Astrophysics, 60 Garden St., Cambridge, MA 02138, USA
\\$^9$Department of Astronomy, Case Western Reserve University, OH, USA
\\$^{10}$Center for Astrophysics and Supercomputing, Swinburne University, Hawthorne, VIC, Australia
\\$^{11}$Institute of Cosmology \& Gravitation, University of Portsmouth, Dennis Sciama Building, Portsmouth PO1 3FX, UK 
\\$^{12}$ Campus of International Excellence UAM+CSIC, Cantoblanco, E-28049 Madrid, Spain
\\$^{13}$ Instituto de F\'{\i}sica Te\'orica, (UAM/CSIC), Universidad Aut\'onoma de Madrid, Cantoblanco, E-28049 Madrid, Spain
\\$^{14}$ Instituto de Astrof\'{\i}sica de Andaluc\'{\i}a (CSIC), Glorieta de la Astronom\'{\i}a, E-18080 Granada, Spain
\\$^{15}$Lawrence Berkeley National Lab, 1 Cyclotron Road, Berkeley, CA 94720, USA
\\$^{16}$Department of Astronomy and Astrophysics, The Pennsylvania State University, University Park, PA 16802
\\${17}$Institute for Gravitation and the Cosmos, The Pennsylvania State University, University Park, PA 16802
\\$^{18}$Steward Observatory, University of Arizona, AZ, USA
\\$^{19}$Department of Physics, University of California Berkeley, CA, USA
\\$^{20}$Department of Astronomy and CCAPP, Ohio State University, Columbus, OH, USA
}

\makeatother

\usepackage{babel}
\begin{document}
\label{firstpage}
\maketitle
\begin{abstract}
We report on the small scale ($0.5<r<40h^{-1}\mathrm{Mpc}$) clustering
of $78895$ massive ($M_{*}\sim10^{11.3}M_{\odot}$) galaxies at $0.2<z<0.4$
from the first two years of data from the Baryon Oscillation Spectroscopic
Survey (BOSS), to be released as part of SDSS Data Release 9 (DR9).
We describe the sample selection, basic properties of the galaxies,
and caveats for working with the data. We calculate the real- and
redshift-space two-point correlation functions of these galaxies,
fit these measurements using Halo Occupation Distribution (HOD) modeling
within dark matter cosmological simulations, and estimate the errors
using mock catalogs. These galaxies lie in massive halos, with a mean
halo mass of $5.2\times10^{13}h^{-1}M_{\odot}$, a large scale bias
of $\sim2.0$, and a satellite fraction of $12\pm2\%$. Thus, these
galaxies occupy halos with average masses in between those of the
higher redshift BOSS CMASS sample and the original SDSS I/II LRG sample. 
\end{abstract}

\section{Introduction}

\label{sec:Introduction}

The large scale structure traced by galaxies is fundamentally dependent
on the cosmology of the early Universe. Because of the growth of early
perturbations due to gravitational attraction, massive dark matter
halos are more strongly clustered than less massive halos. As massive
galaxies preferentially live in massive halos, we can use large surveys
of massive galaxies to probe the evolution of dark matter halos through
cosmic time. Past galaxy redshift surveys such as the Two-Degree Field
Galaxy Redshift Survey \citep{2003astro.ph..6581C} and Sloan Digital
Sky Survey \citep[SDSS;][]{2000AJ....120.1579Y} I/II galaxy samples
\citep{2001AJ....122.2267E,2002AJ....124.1810S} provided large catalogs
to constrain the properties of galaxies and their halos in the relatively
local Universe. The Baryon Oscillation Spectroscopic Survey \citep[BOSS;][]{2009astro2010S.314S},
part of the SDSS-III project \citep{2011AJ....142...72E}, includes
populations of galaxies and quasars to probe the evolution of large
scale structure over cosmic time.

This paper presents the first measurements of the clustering of the
low redshift ($0.2<z<0.4$) BOSS galaxy sample. This study includes
observations from June 2010 through June 2011, and compares this sample
with the high redshift BOSS sample ($0.43<z<0.7$) that was analyzed
by \citet{2011ApJ...728..126W}, and \citet{2012arXiv1203.6594A},
as well as earlier SDSS I/II galaxy samples \citep{2001AJ....122.2267E,2005ApJ...621...22Z,2008MNRAS.387.1045W,2009MNRAS.397.1862P,2009ApJ...707..554Z,2010MNRAS.405.2534T}
from a similar redshift range. The galaxy redshift information used
in this analysis will be released as part of the Data Release 9 (DR9)
public catalog. BOSS will use these data and the CMASS galaxy sample
\citep{2012arXiv1203.6594A} to measure the Baryon Acoustic Oscillation
(BAO) signature in the correlation function and power spectrum to
high precision across a range of redshifts. This information will
provide precise constraints on cosmology that are nearly orthogonal
to those provided by cosmic microwave background \citep[e.g.][]{2003ApJS..148....1B,2011ApJS..192...18K}
and supernova studies \citep[e.g.][]{2008ApJ...686..749K,2009ApJS..185...32K,2009ApJ...703.1374S,2010MNRAS.401.2331L,2010ApJ...716..712A}.

SDSS I/II had two galaxy samples: the main sample \citep{2002AJ....124.1810S},
with a mean redshift $z\sim0.1$, intended to broadly sample all classes
of galaxies over a wide range of luminosity and color, and the Luminous
Red Galaxy (LRG) sample \citep{2001AJ....122.2267E}, with a mean
redshift $z\sim0.3$, intended to provide a large effective volume
for large scale structure studies. The LRG sample provided the first
clear detection of the BAO feature in the galaxy correlation function
and power spectrum \citep{2005ApJ...633..560E}, motivating the design
of BOSS. The BOSS galaxy samples were selected to produce a mostly
sample variance limited measure of BAO to $z=0.7$. BOSS thus includes
a $z<0.45$ sample (LOWZ) with higher number density than the SDSS
I/II LRGs and a higher redshift sample (CMASS) of similar space density.

The primary goal of this paper is to characterize the LOWZ BOSS sample
and compare it with other samples of massive, red galaxies at similar
redshifts. We begin with a description of the galaxy sample, including
selection criteria and caveats in Section \ref{sec:The-Sample}. We
describe the overall clustering properties in Section \ref{sec:Real measurements},
our Halo Occupation Distribution (HOD) model in Section \ref{sec:HOD},
our error derivation in Section \ref{sub:error estimates}, and our
technique for fitting the correlation function in Section \ref{sub:MockCatalogsandMCMC}.
As a test of our fitting procedure, we compare the resulting HOD models
with the measured redshift space clustering in Section \ref{sec:Redshift measurement}.
In Section \ref{sec:CMASS comparison} we compare the properties of
this sample with a number of previous studies. Section \ref{sec:Conclusions}
summarizes our results and some technical details of our HOD modeling
appear in an appendix. For this work, we quote distances as comoving
separations in $h^{-1}\mathrm{Mpc}$ and convert redshifts to distances
assuming a flat $\mbox{\ensuremath{\Lambda}}\mbox{CDM}$ cosmology,
with $\Omega_{M}=0.274$.

\section{The Sample}

\label{sec:The-Sample}

The Sloan Digital Sky Survey \citep[SDSS;][]{2000AJ....120.1579Y}
mapped over one third of the sky using the dedicated 2.5-m Sloan Telescope
\citep{2006AJ....131.2332G} located at Apache Point Observatory in
New Mexico. A drift-scanning mosaic CCD camera \citep{1998AJ....116.3040G}
imaged the sky in five photometric bandpasses \citep{1996AJ....111.1748F,2002AJ....123.2121S,2010AJ....139.1628D}
to a $5\sigma$ limiting magnitude of $r\simeq22.5$. The imaging
data were processed through a series of pipelines that perform astrometric
calibration \citep{2003AJ....125.1559P}, photometric reduction \citep{2001ASPC..238..269L},
and photometric calibration \citep{2008ApJ...674.1217P}. The magnitudes
were corrected for Galactic extinction using the maps of \citet{1998ApJ...500..525S}.
BOSS, as part of the SDSS-III survey \citep{2011AJ....142...72E},
has imaged an additional $2400$ square degrees of the South Galactic
sky in a manner identical to the original SDSS imaging.

BOSS targetted two galaxy samples \citep{2012Padmanabhan.inprep}:
CMASS, at $\bar{z}\sim0.5$ and initially analyzed in \citet{2011ApJ...728..126W},
and LOWZ, at $\bar{z}\sim0.3$ , which is the focus of this study.
This sample was selected as an extension of the SDSS I/II LRG (henceforth:
Legacy) sample \citep{2001AJ....122.2267E}, with three times its
space density. The goal of both BOSS galaxy target selection methods
is to identify luminous, highly biased galaxies, with a galaxy number
density $\bar{{N}}(z)\sim3\times10^{-4}h^{3}\mathrm{Mpc}^{-3}$. These
galaxies should represent the most strongly clustered galaxies at
that space density and redshift range. While the CMASS sample was
targeted to be an approximately mass-limited sample of galaxies with
a range of colors (about $25\%$ are blue), LOWZ consists primarily
of red galaxies. Up to $\sim30\%$ of LOWZ targets were observed during
the Legacy survey, and thus already have a redshift. This reduces
the number of new redshifts required, but slightly complicates the
analysis, as the completeness must be handled differently for Legacy
and new BOSS redshifts. The Legacy redshifts do not impact the uniformity
of the targeting in a given region, as they were ignored when assigning
fibers to targets (see the discussion about completeness in Section
\ref{sec:Real measurements}).

When defining colors, we use the SDSS model magnitudes which were
computed using either an exponential \citep{1970ApJ...160..811F}
or a de Vaucouleurs \citep{1948AnAp...11..247D} light profile fit
to the $r$-band only, and are denoted with the $\mathrm{mod}$ subscript.
Composite model magnitudes are computed using the best-fit linear
combination of an exponential and a de Vaucouleurs light profile fit
to each photometric band independently, and are denoted with the subscript
$\mathrm{cmod}$. Point-spread function (PSF) magnitudes are computed
by fitting a PSF model to the galaxy, and are denoted with the subscript
$\mathrm{psf}$.

\begin{figure*}
\begin{centering}
\includegraphics[width=1\linewidth]{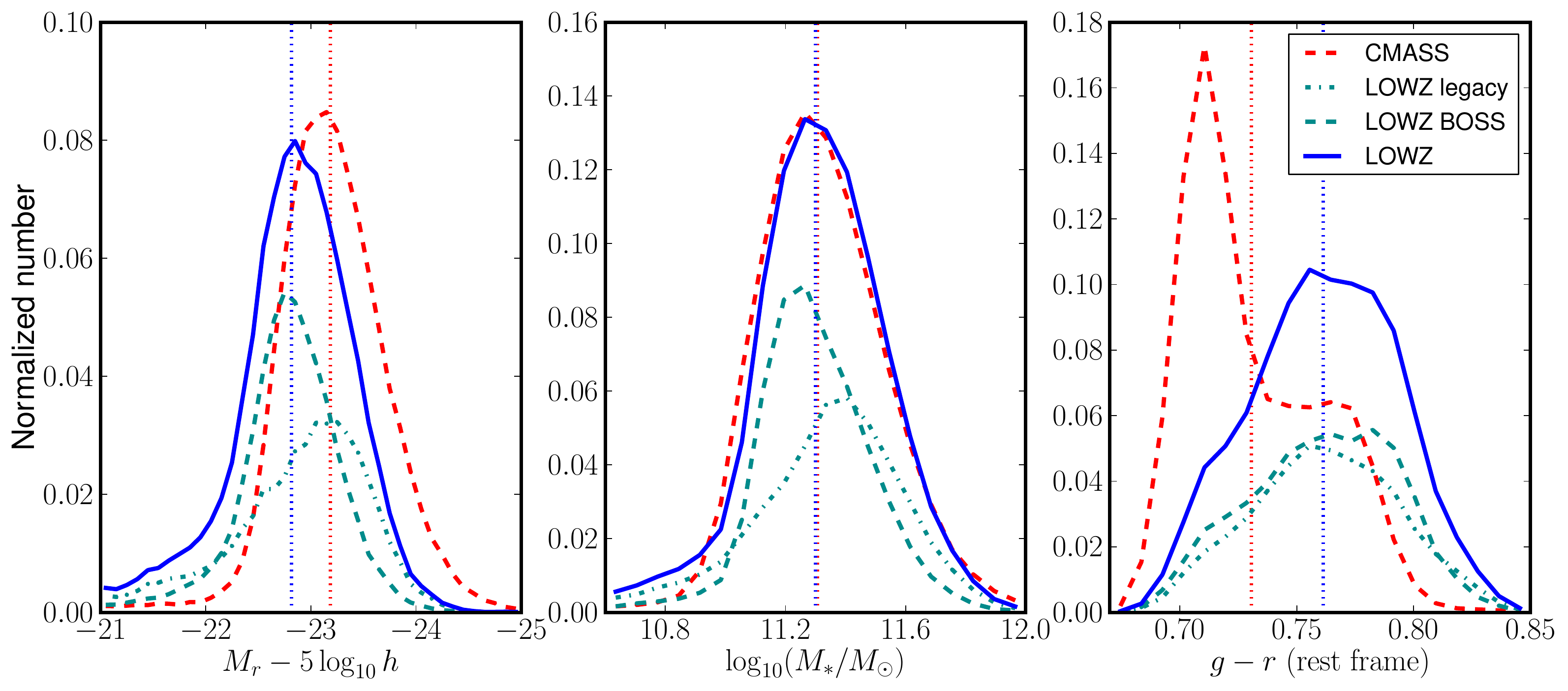}
\par\end{centering}

\caption{\label{fig:absmags}All Panels: LOWZ (solid blue), split into SDSS
I/II LRGs (Legacy, dot-dashed cyan) and new BOSS (dashed cyan) galaxies,
and CMASS (dashed red). Vertical dotted lines show the mean values
for LOWZ and CMASS. All values taken from \citet{2012arXiv1207.6114M}
using LRG models from \citet{2009MNRAS.394L.107M}. Left panel: Absolute
$r$-band rest frame magnitudes. These magnitudes are K+e-corrected
to $z=0$, including corrections for passive evolution. CMASS galaxies
and LOWZ galaxies with redshifts from Legacy are more luminous, in
general, than the new LOWZ BOSS galaxies. Center panel: Stellar mass
in $\log_{10}(M_{*}/M_{\odot})$. BOSS galaxies with Legacy redshifts
have higher masses, on average, than those with new BOSS redshifts.
Right panel: K+e-corrected rest frame $g-r$ colors. CMASS contains
bluer galaxies than LOWZ, while the colors of Legacy and new BOSS
galaxies are very similar.}
\end{figure*}

The LOWZ galaxy target selection algorithm is a straightforward extension
of the method of \citet{2001AJ....122.2267E} to fainter magnitudes
to increase the number density. We define two parameters based on
the $ugriz$ model magnitudes,

\begin{equation}
c_{\parallel}=0.7(g_{\mathrm{mod}}-r_{\mathrm{mod}})+1.2(r_{\mathrm{mod}}-i_{\mathrm{mod}}-0.18)\label{eq:cparallel}
\end{equation}
\begin{equation}
c_{\perp}=(r_{\mathrm{mod}}-i_{\mathrm{mod}})-(g_{\mathrm{mod}}-r_{\mathrm{mod}})/4.0-0.18.\label{eq:cperp}
\end{equation}
We target galaxies that are luminous and red with a redshift $z\lesssim0.4$
with the following cuts.
\begin{eqnarray}
r_{\mathrm{cmod}} & < & 13.5+c_{\parallel}/0.3\label{eq:r vs. cpll cut}\\
|c_{\perp}| & < & 0.2\label{eq:cperp cut}\\
16 & <r< & 19.6\label{eq:rmod cut}\\
r_{\mathrm{psf}}-r_{\mathrm{cmod}} & > & 0.3\label{eq:star/gal}
\end{eqnarray}
This selection follows the color tracks of a passively evolving stellar
population (Eq. \ref{eq:cperp}), and selects an approximately absolute
magnitude limited sample (Eq. \ref{eq:r vs. cpll cut}) with a sliding
cut in color and luminosity. Equation \ref{eq:star/gal} is the primary
star/galaxy separation, based on the difference between a modeled
galaxy light profile, and a PSF profile. Note that this is different
from the LOWZ target selection algorithm used during the first 9 months
of BOSS (roughly, data taken through June 2010), which was affected
by a change due to a bug in the star-galaxy separation. The earlier
data have a lower on-sky density and cannot be corrected to reflect
the new targeting, as was done in \citet{2011ApJ...728..126W} and
\citet{2012arXiv1203.6594A} for CMASS. Because of this issue, we
restrict ourselves to regions that were tiled for spectroscopy with
the corrected target selection, via 
\[
\mathrm{TILE}\geqq10324.
\]
Plates outside this range were drilled with the earlier target selection
and star/galaxy separation algorithm. We recommend users of this sample
who wish to perform large scale structure analyses on the LOWZ sample
apply the same cuts on the catalog data to remove the early data with
a different selection.

We define ``good'' redshifts as follows, where ``\&\&'' is the
bitwise \emph{and} operator. For new BOSS observations, we require
\begin{itemize}
\item (BOSS\_TARGET1 \&\& $2^{0}$) > 0
\item SPECPRIMARY == 1
\item ZWARNING\_NOQSO == 0
\end{itemize}
while for SDSS Legacy observations we require
\begin{itemize}
\item SPECPRIMARY == 1
\item ZWARNING == 0
\end{itemize}
These parameters are part of the SDSS DR9 catalog: BOSS\_TARGET1 is
a bitmask containing the target selection flags, SPECPRIMARY identifies
the best spectrum among multiple observations, and ZWARNING\_NOQSO
and ZWARNING are bitmasks listing potential problems with the redshift
fit, with a value of 0 representing no obvious problems. We do not
require that the spectrum is identified as a galaxy spectrum. i.e.,
if an object passed the targeting cuts, and the spectrum satisfied
the above requirements, a non-galaxy spectrum (e.g. strong emission
lines, quasars) would be included in the sample.

In Figure \ref{fig:absmags} we show the absolute $r$-band rest frame
magnitude, stellar mass, and $g-r$ rest frame color distributions
of this sample, K+e-corrected to $z=0$, using values from \citet{2012arXiv1207.6114M}.
The stellar masses were calculated assuming a \citet{2001MNRAS.322..231K}
initial mass function, including stellar mass loss due to stellar
evolution. We also plot distributions for the CMASS sample and split
the LOWZ sample into Legacy and new BOSS objects to show the difference
between SDSS I/II targets and new BOSS targets. LOWZ galaxies are
typically half a magnitude fainter than CMASS galaxies, but CMASS
contains more blue galaxies. LOWZ galaxies which were observed as
part of SDSS I/II are typically brighter and have $\sim0.4\,\mbox{dex}$
higher stellar masses, which is to be expected given the brighter
magnitude selection imposed as part of that survey. LOWZ galaxies
have very similar colors to Legacy galaxies, and are typically redder
than CMASS galaxies. 

\begin{figure}
\begin{centering}
\includegraphics[width=1\columnwidth]{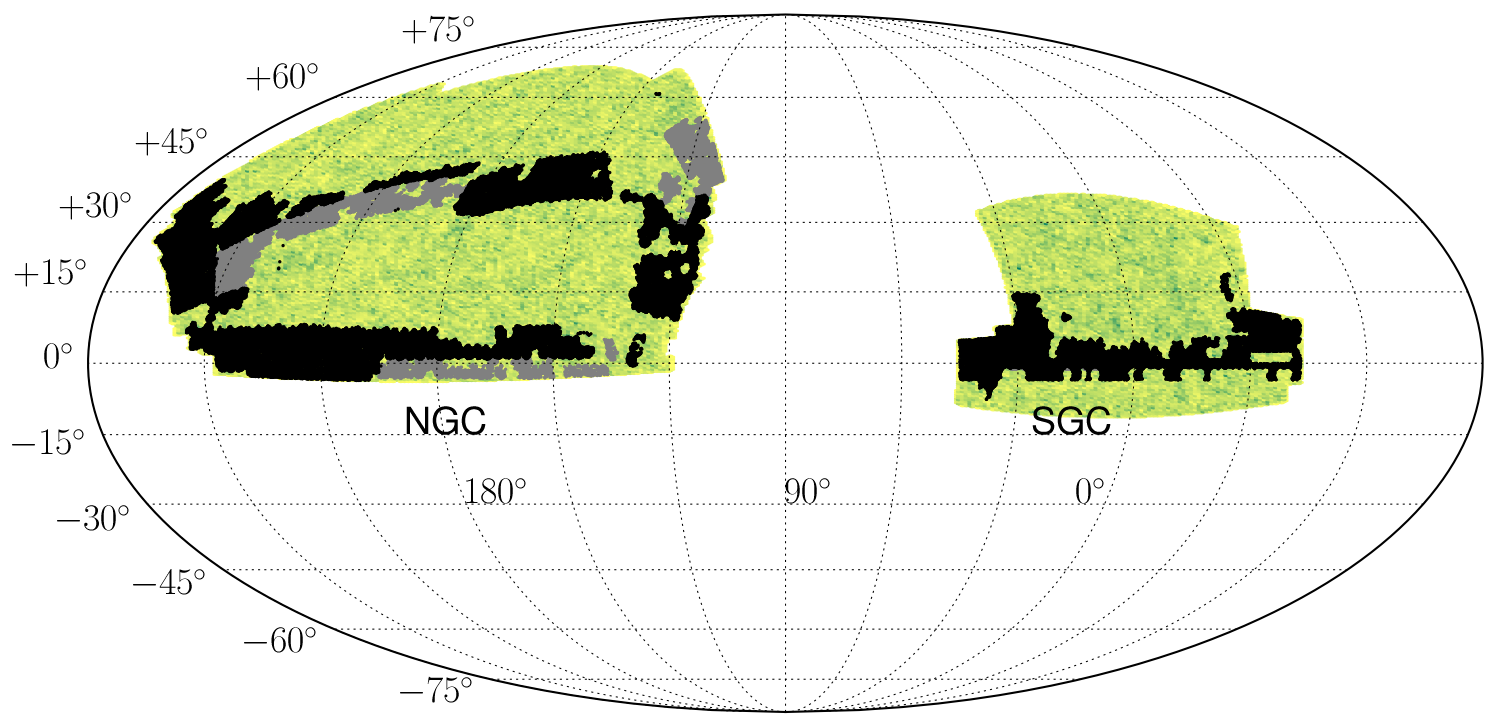}
\par\end{centering}

\caption{\label{fig:on-sky-distribution}The on-sky distribution (equatorial
coordinates) of our sample, with the north and south Galactic caps
(NGC and SGC, respectively) labeled for reference. The black regions
show the data used in the current study, while the light grey areas
are the regions that had to be dropped because of the change in target
selection. The background yellow/green color field shows the on-sky
density of LOWZ targets in the full BOSS target area, representing
the total area that will be covered when the survey is completed in
2014. }
\end{figure}

Figure \ref{fig:on-sky-distribution} displays the on-sky distribution
of the current sample. As noted above, because of changes in target
selection we are not able to use roughly the first year's worth of
data. This reduces the total sky coverage compared with the DR9 CMASS
sample of \citet{2012arXiv1203.6594A} by $1205$ square degrees ($0.367$
steradians), and removes a large part of the area studied in \citet{2011ApJ...728..126W}.
The area with good LOWZ data shown covers $2467$ square degrees ($0.7517$
steradians) on the sky.

We consider data from the North Galactic Cap (NGC) and South Galactic
Cap (SGC) separately in our analysis for a number of reasons. We lack
a dark matter simulation that resolves the host halos of BOSS galaxies
and is large enough to fit the NGC and SGC in a single simulation
box. The SGC has a $\sim8\%$ higher target density than the NGC,
and thus a higher galaxy number density, mostly due to differences
in photometric calibration and reddening correction between the hemispheres.
\citet{2010ApJ...725.1175S} and \citet{2011ApJ...737..103S} found
 a difference in the SDSS colors between the NGC and SGC, resulting
in a $0.015\,\mathrm{mag}$ offset in $c_{\parallel}$ (eq. \ref{eq:cparallel}
and \ref{eq:r vs. cpll cut}). The ``ubercal'' SDSS photometry\citep{2008ApJ...674.1217P}
uses overlapping regions of images to cross-calibrate the photometric
measurments. As the NGC and SGC are not contiguous they are less well
cross-calibrated than they are internally calibrated. This produces
some of the measured color difference between the NGC and SGC. \citet{2011ApJ...737..103S}
also identified a slight systematic error in the reddening correction
that was applied before targetting. \citet{2011MNRAS.417.1350R} describe
the use of the \citet{2011ApJ...737..103S} color offsets, determined
from stellar spectra, to correct for the north/south asymmetry. They
find that the NGC/SGC CMASS and LOWZ number density differences are
completely consistent with the level of color offset found by \citet{2011ApJ...737..103S},
but note the inherent uncertainties in the offsets and resulting corrections.
They also find that these, and other, systematics more strongly affect
the CMASS sample than the LOWZ sample, and are most important for
clustering studies on the largest scales. As we will show, the resulting
two LOWZ galaxy populations are not significantly different in their
clustering properties (see results in Sections \ref{sec:HOD}, \ref{sec:Redshift measurement},
and \ref{sec:CMASS comparison}).  We provide values for the NGC and
SGC separately, and also provide minimum-variance weighted values
for NGC+SGC (hereafter, Full) when appropriate.

\begin{figure}
\begin{centering}
\includegraphics[width=1\columnwidth]{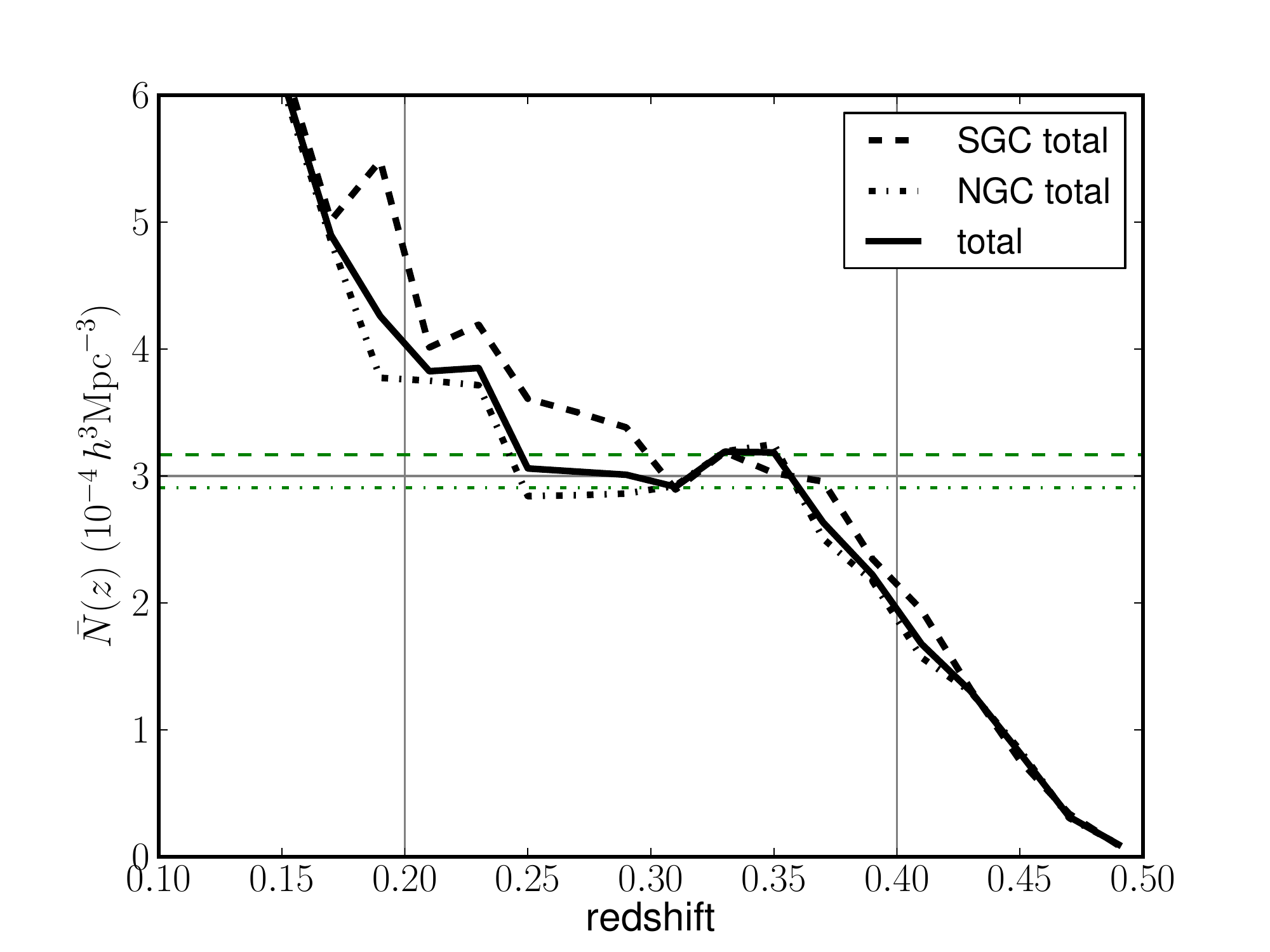}
\par\end{centering}

\caption{\label{fig:The-redshift-distribution}The redshift distribution of
the BOSS LOWZ sample. The SGC and NGC are plotted separately as dashed
and dashed-dotted lines, respectively; showing the higher number density
in the SGC. The solid black line gives the total $\bar{{N}}(z)$ distribution
for SGC+NGC. Our redshift range of $0.2<z<0.4$ is marked with thin
vertical lines. The original survey galaxy density goal of $\bar{{N}(z)}=3\times10^{-4}h^{3}\mathrm{Mpc}^{-3}$
is shown for reference (thin horizontal line), while the green horizontal
lines show the NGC and SGC effective mean density over the $0.2<z<0.4$
range (values given in Table \ref{tab:galaxy-statistics})}
\end{figure}

\begin{table}
\caption{\label{tab:galaxy-statistics}Statistics of the $0.2<z<0.4$ galaxy
sample}

\begin{centering}
\begin{tabular}{ccc}
\hline 
\hline region & $\mbox{N}_{galaxy}$ & $\bar{N}(z)\,(10^{-4}h^{3}\mathrm{Mpc}^{-3}$)\tabularnewline
\hline 
SGC (Legacy) & 3946 & 1.502\tabularnewline
SGC (BOSS) & 19558 & 2.681\tabularnewline
SGC (Legacy+BOSS) & 23504 & 3.167{*}\tabularnewline
\hline 
NGC (Legacy) & 18332 & 0.990\tabularnewline
NGC (BOSS) & 37059 & 2.005\tabularnewline
NGC (Legacy+BOSS) & 55391 & 2.907{*}\tabularnewline
\hline 
Full (Legacy) & 22278 & 1.053\tabularnewline
Full( BOSS) & 56617 & 2.198\tabularnewline
Full (Legacy+BOSS) & 78895 & 2.981\tabularnewline
\hline 
\end{tabular}
\par\end{centering}

\centering{}{\footnotesize The $\bar{{N}}(z)$ values marked with
an asterisk are used in the MCMC fitting procedure.}
\end{table}

Figure \ref{fig:The-redshift-distribution} presents the galaxy number
density of our samples as a function of redshift. The vertical lines
denote the redshift range used in this work ($0.2<z<0.4$). We restrict
to this range as it provides a relatively uniform number density across
the redshift interval, and to distinguish it from the CMASS sample
studied by \citet{2011ApJ...728..126W}, which was restricted to $0.4<z<0.7$.
The dramatic increase in the number density below $z\sim0.2$ is due
in part to more than just massive red galaxies falling into the target
selection. Additionally, \citet{2011MNRAS.417.1114T} and \citet{2012arXiv1202.6241T}
found that SDSS II LRGs below $z\lesssim0.2$ had different dynamical
growth than LRGs at higher redshifts. A more in-depth study of the
uniformity and completeness of the full LOWZ sample is in progress.
The difference between the NGC (dash-dotted lines) and SGC (dashed
lines) number density is clearly visible in the plot. We present some
basic statistics of our sample in Table \ref{tab:galaxy-statistics},
including separate values for the SGC, NGC, and Full NGC+SGC, and
Legacy, BOSS, and Legacy+BOSS redshift samples. One can see the significantly
larger number of new BOSS redshifts compared to Legacy in the SGC,
as SDSS I/II only observed three stripes in that region covering about
$700$ square degrees, compared with $3100$ square degrees of imaging
available to BOSS \citep{2011ApJS..193...29A}. The BOSS redshift
densities in the two hemispheres should not be directly compared,
because of the different number of Legacy redshifts in the NGC vs.
SGC.

\section{Real Space Clustering}

\label{sec:Real}

\subsection{Measurements}

\label{sec:Real measurements}

The correlation function $\xi(r)$ \citep{1980lssu.book.....P} measures
the excess probability of finding a galaxy in a volume element, $dV$,
at separation $r$ from a randomly selected galaxy,
\begin{equation}
dP(r)=N_{G}(1+\xi(r))dV,\label{eq:CF-probability}
\end{equation}
where $N_{G}$ is the mean galaxy number density. We use the estimator
of \citet{1993ApJ...412...64L},
\begin{equation}
\xi(r)=\frac{DD-2DR+RR}{RR}.\label{eq:LS-estimator}
\end{equation}
Compared to the ``natural'' estimator $DD/RR-1$, the Landy and
Szalay estimator reduces geometrical edge effects and minimizes variance.

\begin{figure*}
\begin{centering}
\includegraphics[width=1\linewidth]{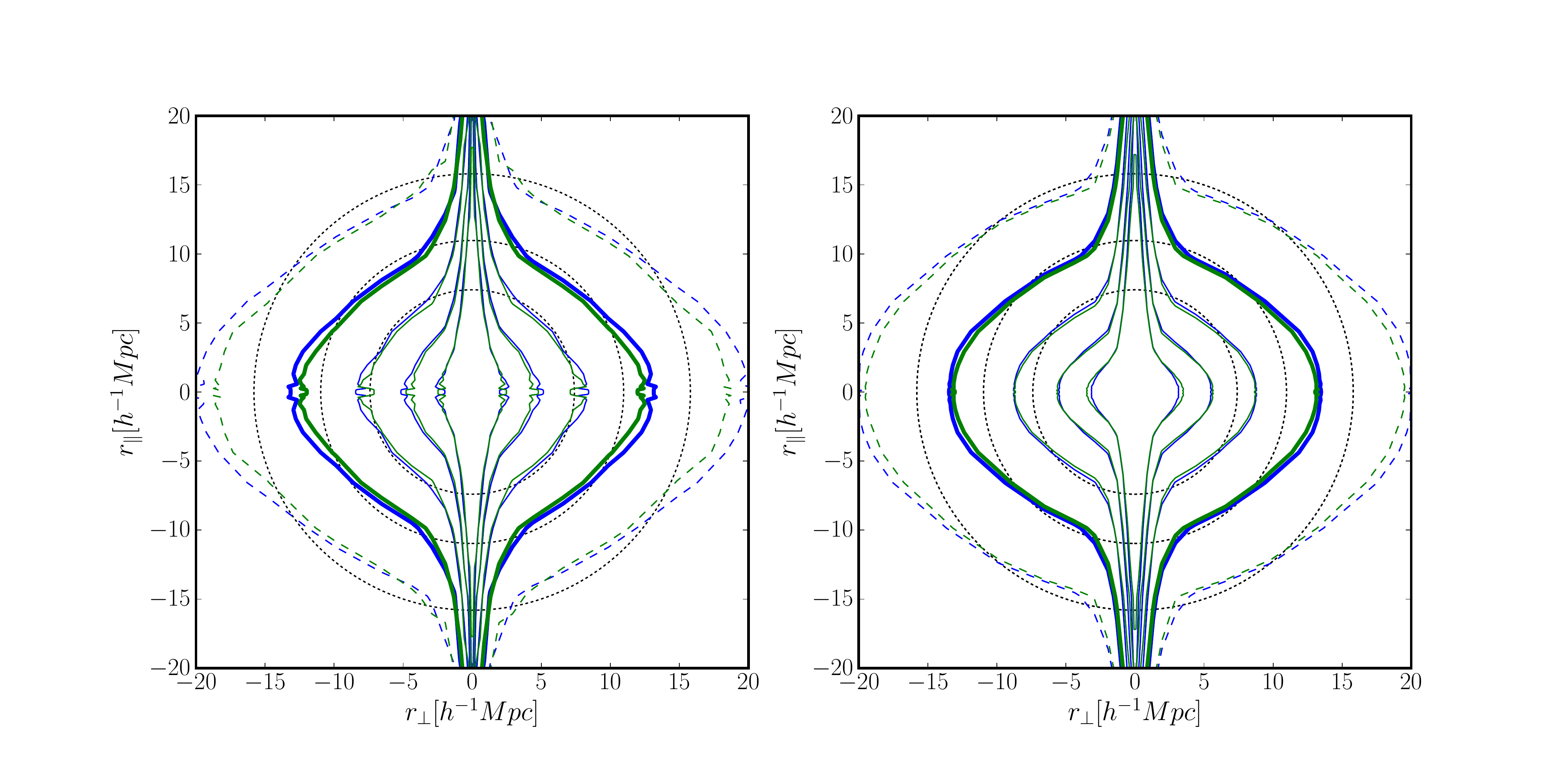}
\par\end{centering}

\caption{\label{fig:projected-correlation-contours}Left panel: contours of
the correlation function $\xi(r_{\perp},r_{\parallel})$ (smoothed
with a Gaussian kernel of radius $0.5\, h^{-1}\mathrm{Mpc}$ for clarity)
of the components parallel ($r_{\parallel}$) and perpendicular ($r_{\perp}$)
to the line of sight, for BOSS LOWZ galaxies with redshifts in the
range $0.2<z<0.4$. In both plots the blue lines show results for
NGC data and green lines show results for SGC data. Solid contours
are shown at $\xi=(1,2,4,8)$, dashed contours at $\xi=(0.5)$, and
a thicker solid line is shown for $\xi=1$. Thin black dotted circles
are plotted for $\xi(r)=(2,1,0.5)$ for reference. Right panel: the
same, but for our mock galaxy catalog (Section \ref{sub:error estimates}).
Note that similar features (e.g. finger-of-god) are present in the
mocks. In both panels, the discontinuity at $r_{\parallel}=0$ is
due to small number fluctuations in the smallest bins.}
\end{figure*}

We use this estimator to calculate a two-point galaxy correlation
function of pair separations parallel ($r_{\parallel}$) and perpendicular
($r_{\perp}$) to the line of sight, $\xi(r_{\parallel},r_{\perp})$.
We then compute the projected correlation function, $\omega_{p}(r_{\perp})$,
which reduces the effect of redshift-space distortions, by integrating,

\begin{equation}
\omega_{p}(r_{\perp})=2\int_{0}^{r_{\parallel,\mathrm{max}}}\xi(r_{\parallel},r_{\perp})dr_{\parallel},\label{eq:wp_from_xi(rperp,rpll)}
\end{equation}
where we take $r_{\parallel,\mathrm{max}}=75h^{-1}\mathrm{Mpc}$ as
the upper limit of the integral. The integral is stable around this
value, while not contributing noise from large $r_{\parallel}$ bins.
The relationship between the projected correlation function, $\omega_{p}(r_{\perp})$
, and the real-space correlation function is \citep[e.g.][]{1983ApJ...267..465D}
\begin{equation}
\omega_{p}(r_{\perp})=2\int_{0}^{y_{\mathrm{max}}}\xi[(r_{\perp}^{2}+y^{2})^{1/2}]dy\label{eq:wp_from_xi(r)}
\end{equation}
\label{sub:HOD}which is used in our full-box mock simulations to
more quickly compute $w_{p}(r_{\perp})$ (see Section \ref{sec:HOD}).
We again integrate to $y_{\mathrm{max}}=75h^{-1}\mathrm{Mpc}$ as
a balance between including most of the information from the correlation
function, and limiting noise from large radius bins. We tested and
confirmed that this integral produces almost identical results to
the integral over the correlation function in Eq. \ref{eq:wp_from_xi(rperp,rpll)},
as long as $\xi(r)$ is measured in small enough bins to allow for
smooth interpolation.

Not all galaxy targets in each region on the sky were assigned spectroscopic
fibers. Specifically, the completeness of the survey varies between
the different regions defined by \emph{sectors} which consist of disjoint
regions defined by the overlap of spectroscopic tiles. In addition,
as mentioned above, some of our targets come ``pre-observed'', i.e.,
with redshifts from SDSS I/II. We thus separate galaxies into two
groups and calculate their completeness\emph{,} $\mbox{fgot}$, on
a per-sector basis as\label{eq:completeness}
\begin{eqnarray}
\mathrm{fgot}_{\mathrm{BOSS}} & = & N_{\mathrm{BOSS}}/(N_{\mathrm{targets}}-N_{\mathrm{Legacy}})\label{eq:completeness-1}\\
\mathrm{fgot}_{\mathrm{Legacy}} & = & 1\nonumber 
\end{eqnarray}
where $N_{\mathrm{BOSS}}$ is the number of new, not confirmed as
star, BOSS redshifts, $N_{\mathrm{targets}}$ is the number of LOWZ
targets and $N_{\mathrm{Legacy}}$ is the number of previously acquired
Legacy redshifts that were targeted per the algorithm described above.
For Legacy redshifts the completeness is defined to be 1, as they
can be considered a separate, fixed, sample whose redshifts are pre-determined.
Known Legacy redshifts were excluded during targeting, so they do
not affect whether any BOSS targets were allocated a fiber. If $N_{\mathrm{targets}}-N_{\mathrm{Legacy}}=0$,
the sector is assigned a completeness of $1$, as all of the targets
have Legacy redshifts, and no new BOSS redshifts were required.

When computing the correlation function, we weight galaxies by $\mathrm{fgot}^{-1}$
(see Eq. \ref{eq:completeness-1}), restricted to only those galaxies
which lie in sectors with $\mathrm{fgot\mbox{}}>0.5$. For calculating
the correlation function, we generate random points uniformly in all
regions with $\mathrm{fgot}>0.5$, and assign all randoms a weight
of $1$. We generate $100$ times the number of data points for our
random catalog as we have observations, ensuring the variance of the
results is not affected by the random catalog. We assign redshifts
to the randoms by smoothing the NGC and SGC redshift distributions
with a 7th order Chebyshev polynomial and drawing from each of the
resulting distributions separately for the NGC and SGC random catalogs,
respectively.

To correct for fiber-collisions, we use the nearest-neighbor redshift
method \citep{2002ApJ...571..172Z,2005ApJ...621...22Z,2006ApJS..167....1B}.
To each galaxy within a ``collision group'' that does not have a
good redshift, we assign the redshift of the nearest galaxy within
$62\arcsec$. Although this method is known to over-correct below
the fiber-collision radius ($62\arcsec$\textbf{ }for BOSS, corresponding
to $0.235h^{-1}\mathrm{Mpc}$ at the highest redshift in our sample),
it is a nearly exact correction at radii larger than twice the fiber-collision
radius. \citet{2011arXiv1111.6598G} demonstrated a more exact method
for corrections below the fiber-collision radius, but their method
reduces the number of points included in the correlation function
calculation, and thus increases the variance of the measurement. As
the smallest radius bin we consider is well above the fiber-collision
radius, and the results of \citet{2011arXiv1111.6598G} show close
correspondence between their new method and the nearest neighbor method
for large radii, we will adopt the simpler method in this work.

\begin{figure*}
\begin{centering}
\includegraphics[width=0.5\linewidth]{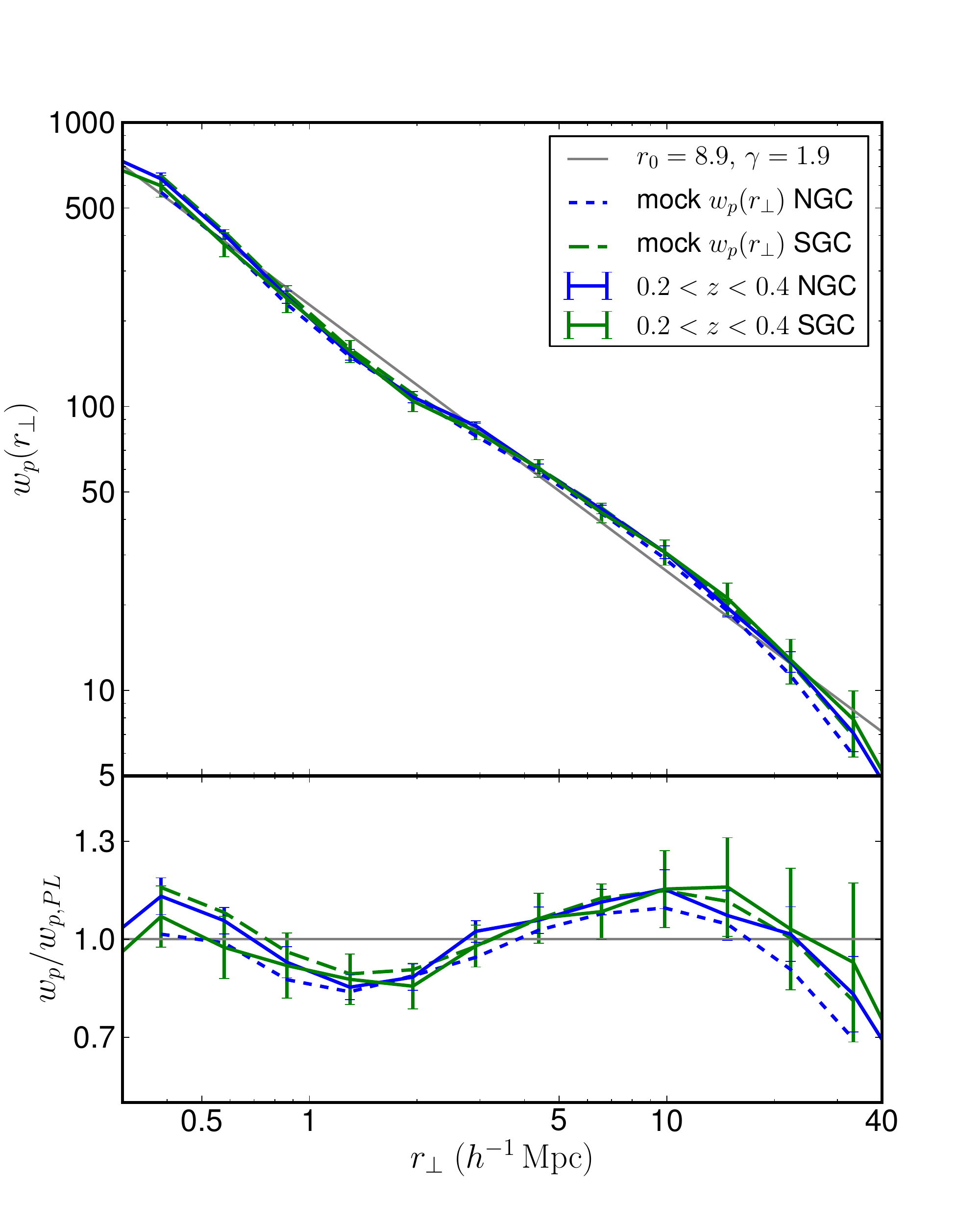}\includegraphics[width=0.5\linewidth]{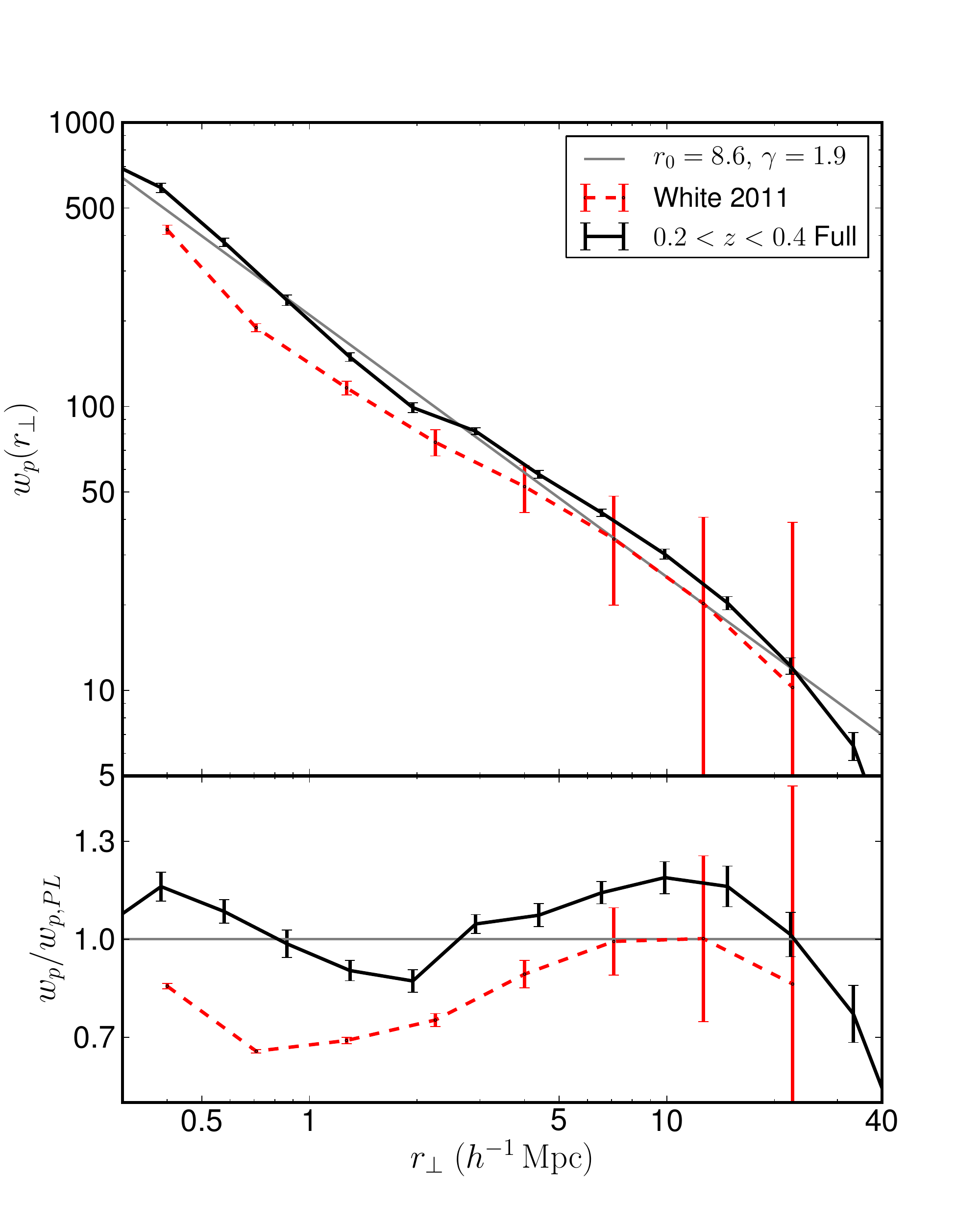}
\par\end{centering}

\caption{\label{fig:projected-correlation}Left: NGC and SGC, data and mocks.
Right: NGC+SGC Full sample. In each plot, the upper panel shows the
projected correlation function for BOSS LOWZ galaxies with redshifts
in the range $0.2<z<0.4$. Dashed lines show the mean (not the single
best-fitting model) mock correlation functions from our MCMC fitting
procedure. Errors are computed from the square-root of the diagonal
of our covariance matrix, generated using the mean HOD parameters.
The thin line shows a simple best-fit power law, $\xi_{PL}=\left(\frac{r}{r_{0}}\right){}^{-\gamma}$,
transformed to $w_{p,PL}(r_{\perp})$ using eq. \ref{eq:wp_from_xi(r)},
to guide the eye. Also plotted on the right is the mean CMASS correlation
function from \citet{2011ApJ...728..126W} for comparison. The lower
panels show the same curves as above, but with each curve divided
by the power law fit to emphasize the differences between the various
correlation functions, and to show that a pure power law fit is not
appropriate for this data. The NGC and SGC correlation functions match
within the errors. The correlation functions clearly display the inflection
at $1-2h^{-1}\mathrm{Mpc}$ that marks the transition between the
one-halo and two-halo regimes \citep{2004ApJ...608...16Z}.}
\end{figure*}

We show the correlation function contours in $(r_{\parallel},r_{\perp})$
in Figure \ref{fig:projected-correlation-contours}, alongside the
mock catalog 2D correlation function for comparison. The mocks show
similar redshift space distortions to the data, with Fingers-of-God
and large scale squashing due to the Kaiser effect \citep{1987MNRAS.227....1K}
clearly evident. Our procedure for generating the redshift space mock
catalogs is described in Section \ref{sub:error estimates}. Following
Eq. \ref{eq:wp_from_xi(rperp,rpll)}, we integrate this 2D correlation
function to get the projected correlation function, $w_{p}(r_{\perp})$
shown inFigure \ref{fig:projected-correlation}. The errors are estimated
from our mock catalogs (see Section \ref{sec:HOD}), shifted onto
the sky and masked with the coverage map. As noted in Section \ref{sec:The-Sample},
we present the NGC and SGC separately in the left-hand figure, and
the minimum-variance weighted Full NGC+SGC in the right-hand figure.
It is clear from the figure that over the scales considered in this
paper, the clustering of the NGC and SGC are the same within the errors,
differing by less than $1\sigma$. The CMASS errors from \citet{2011ApJ...728..126W}on
large scales are significantly larger than ours, because that analysis
was performed on a much smaller data set ($580\mbox{deg}^{2}$, or
roughly $30\%$ the solid angle). The large scale errors are determined
primarily by the volume of the survey, so our increased volume improves
on those errors. In addition, the geometry of the \citet{2011ApJ...728..126W}
sample naturally split into three disjoint regions, A, B, and C, reducing
the number of large-scale pairs.

\textbf{}

\begin{table*}
\caption{\label{tab:correlationfunction}The projected correlation function,
$w_{p}(r_{\perp})$, in 12 equally spaced bins in $\log_{1.5}(r_{\perp})$.}

\begin{centering}
\begin{tabular}{ccccccccccccc}
\hline 
\hline$r_{\perp}$ & 0.385 & 0.577 & 0.865 & 1.299 & 1.945 & 2.921 & 4.381 & 6.572 & 9.858 & 14.78 & 22.18 & 33.27\tabularnewline
\hline 
Full $w_{p}(r_{\perp})$ & 619.29 & 390.51 & 242.17 & 153.87 & 106.83 & 84.18 & 60.42 & 43.19 & 30.70 & 19.83 & 12.66 & 7.22\tabularnewline
Full $\sigma$ & 27.20 & 18.80 & 9.51 & 5.60 & 4.60 & 3.17 & 2.30 & 1.76 & 1.69 & 1.36 & 0.85 & 0.87\tabularnewline
\hline 
NGC $w_{p}(r_{\perp})$ & 632.95 & 404.09 & 242.84 & 152.32 & 107.88 & 85.42 & 60.33 & 43.41 & 30.70 & 19.52 & 12.63 & 7.07\tabularnewline
SGC $w_{p}(r_{\perp})$ & 598.26 & 372.80 & 240.20 & 156.64 & 104.49 & 81.68 & 60.67 & 42.25 & 30.72 & 21.11 & 12.83 & 7.90\tabularnewline
\hline 
NGC $\sigma$ & 31.49 & 15.53 & 12.40 & 6.735 & 4.888 & 2.746 & 2.292 & 1.527 & 1.584 & 1.376 & 1.042 & 0.9794\tabularnewline
SGC $\sigma$ & 52.58 & 36.36 & 25.99 & 13.78 & 8.555 & 5.376 & 4.354 & 3.322 & 3.138 & 2.755 & 2.309 & 2.070\tabularnewline
\hline 
\hline0.385 & 1 & 0.5380 & 0.5110 & 0.4830 & 0.2790 & 0.2380 & 0.3510 & 0.3610 & 0.4470 & 0.3110 & 0.2710 & 0.2040\tabularnewline
0.577 & \ldots{} & 1 & 0.5150 & 0.4910 & 0.3680 & 0.4150 & 0.4730 & 0.3860 & 0.2860 & 0.2860 & 0.2470 & 0.2990\tabularnewline
0.865 & \ldots{} & \ldots{} & 1 & 0.7090 & 0.6460 & 0.4640 & 0.3520 & 0.3910 & 0.3420 & 0.3250 & 0.1880 & 0.1430\tabularnewline
1.299 & \ldots{} & \ldots{} & \ldots{} & 1 & 0.6150 & 0.5100 & 0.4710 & 0.5010 & 0.5690 & 0.4310 & 0.2470 & 0.0740\tabularnewline
1.945 & \ldots{} & \ldots{} & \ldots{} & \ldots{} & 1 & 0.6180 & 0.6120 & 0.6850 & 0.5140 & 0.4520 & 0.4120 & 0.2920\tabularnewline
2.921 & \ldots{} & \ldots{} & \ldots{} & \ldots{} & \ldots{} & 1 & 0.6570 & 0.5130 & 0.4550 & 0.3870 & 0.2620 & 0.1710\tabularnewline
4.381 & \ldots{} & \ldots{} & \ldots{} & \ldots{} & \ldots{} & \ldots{} & 1 & 0.7850 & 0.7230 & 0.6470 & 0.5540 & 0.3980\tabularnewline
6.572 & \ldots{} & \ldots{} & \ldots{} & \ldots{} & \ldots{} & \ldots{} & \ldots{} & 1 & 0.7550 & 0.6200 & 0.4710 & 0.2550\tabularnewline
9.858 & \ldots{} & \ldots{} & \ldots{} & \ldots{} & \ldots{} & \ldots{} & \ldots{} & \ldots{} & 1 & 0.8300 & 0.5600 & 0.2640\tabularnewline
14.78 & \ldots{} & \ldots{} & \ldots{} & \ldots{} & \ldots{} & \ldots{} & \ldots{} & \ldots{} & \ldots{} & 1 & 0.7980 & 0.5350\tabularnewline
22.18 & \ldots{} & \ldots{} & \ldots{} & \ldots{} & \ldots{} & \ldots{} & \ldots{} & \ldots{} & \ldots{} & \ldots{} & 1 & 0.8750\tabularnewline
33.27 & \ldots{} & \ldots{} & \ldots{} & \ldots{} & \ldots{} & \ldots{} & \ldots{} & \ldots{} & \ldots{} & \ldots{} & \ldots{} & 1\tabularnewline
\hline 
\end{tabular}
\par\end{centering}

\centering{}\textbf{Note:} All bin values are in $h^{-1}\mathrm{Mpc}$,
at the volume-weighted bin center:$((r_{\mathrm{top}}^{3}+r_{\mathrm{bottom}}^{3})/2)^{1/3}$.
The lower part of the table lists the upper triangle of the covariance
matrix as $C_{ij}/(\sigma_{i}\sigma_{j})$.
\end{table*}

We provide the correlation function and error estimates in Table \ref{tab:correlationfunction}
for the NGC, SGC and Full NGC+SGC samples. The values for the Full
sample are computed from a minimum-variance weighted combination of
the NGC and SGC. The bins shown are those used during the fitting
procedure described below. In the same table, we show the covariance
matrix for the Full NGC+SGC data set. The galaxy and random catalogs
used in this analysis are available on the SDSS III website%
\footnote{\url{http://www.sdss3.org/dr9/data_access/vac.php} under Large Scale
Structure Galaxy Catalogs%
}.

\subsection{Halo Occupation Distribution}

\label{sec:HOD}

We estimate the errors in the sample and determine the dark matter
halo statistics using 20 dark matter simulations. These are the same
simulations used in the CMASS analysis of \citet{2011ApJ...728..126W},
with $1500^{3}$ particles of mass $7.6\times10^{10}h^{-1}M_{\odot}$
in a periodic cube $1500h^{-1}\mathrm{Mpc}$ on a side. All 20 simulations
have the same cosmological parameters: $\Omega_{M}=0.274$, $\Omega_{\Lambda}=0.726$,
$w=-1.00$, $\Omega_{b}=0.0457$, $h=0.70$, $n=0.95$, $\sigma_{8}=0.8$,
the same as the parameters given in the introduction for our redshift
to distance conversion. We identify dark matter halos using a friends-of-friends
(FoF) algorithm \citep{1985ApJ...292..371D}, with a linking length
of 0.168. This results in a minimum resolved halo mass for our redshift
slice ($z=0.30$) of $10^{11.88}h^{-1}M_{\odot}$. For more details
on these simulations, see the appendix of \citet{2011ApJ...728..126W}.
Note that we are not fitting cosmological parameters, and our CMASS
comparison is straightforward because the assumed cosmologies are
exactly the same.

We do not have a single cosmological simulation with a sufficiently
large volume and a high enough number density to embed the NGC and
SGC in the same box, but our simulations are large enough if the NGC
and SGC are considered separately. We can safely ignore correlations
between the NGC and SGC in this analysis, as the shortest distance
between an NGC and an SGC galaxy is $53{}^{\circ}$--corresponding
to $>600\, h^{-1}\mathrm{Mpc}$ at the sample's minimum redshift--while
this paper only considers separations below $40\, h^{-1}\mathrm{Mpc}$.
Our technique of fitting each separate region into our simulation
cubes to generate mock galaxy catalogs for error estimation is discussed
in Section \ref{sub:error estimates}.

For an accurate interpretation of galaxy clustering, we use the halo
occupation distribution (HOD) formulation \citep{2000MNRAS.318.1144P,2000MNRAS.318..203S,2000MNRAS.316..107B,2001ApJ...550L.129W,2001ApJ...546...20S,2002ApJ...575..587B}.
The HOD gives the conditional probability that a halo with virialized
mass $M_{\mathrm{halo}}$ contains $N$ galaxies of a particular class.
In this paper, we distinguish central and satellite galaxies, and
we require that halos with a satellite must have a central galaxy.

We determine the number of satellite and central galaxies in each
halo following a halo prescription based on that of \citet{2005ApJ...633..791Z}
with a central galaxy probability of 
\begin{equation}
N_{\mathrm{cen}}(M_{\mathrm{halo}})=\frac{1}{2}\mathrm{erfc}\left[\frac{\ln(M_{cut}/M_{\mathrm{halo}})}{\sqrt{2}\sigma}\right],\label{eq:Ncen}
\end{equation}
and an expected number of satellites equal to

\begin{equation}
N_{\mathrm{sat}}(M_{\mathrm{halo}})=\left(\frac{M_{\mathrm{halo}}-\kappa M_{cut}}{M_{1}}\right){}^{\alpha},\label{eq:Nsat}
\end{equation}
where $M_{cut}$, $M_{1}$, $\sigma$, $\kappa$, and $\alpha$ are
the free-parameters to be fit in our model, described in detail in
Appendix \ref{sec:Appendix-HOD parameter systematics}. Briefly, $M_{cut}$
is a minimum mass for halos to host our galaxies, $M_{1}$ is a typical
mass for halos to host one satellite, $\sigma$ is the scatter between
$M_{*}$ and $M_{\mathrm{halo}}$, $\kappa$ allows the threshold
mass for satelites and centrals to differ, and $\alpha$ is the mass
dependence of the efficiency of galaxy formation. This halo prescription
was created to reproduce the observed luminosity-dependent clustering
and number densities from the SDSS main galaxy \citep{2002AJ....124.1810S}
and LRG samples \citep{2001AJ....122.2267E}. We assume that for a
halo to contain satellites it must first contain a central, so we
only assign satellites to halos with a central galaxy. Thus, the total
expected number of galaxies in a halo of mass $M_{\mbox{halo}}$ is
\begin{eqnarray}
\langle N_{\mathrm{gal}}(M_{\mathrm{halo}})\rangle & = & \langle N_{\mathrm{cen}}(M_{\mathrm{halo}})\rangle(1+\langle N_{\mathrm{sat}}(M_{\mathrm{halo}})\rangle).\label{eq:total-gals}
\end{eqnarray}
We assign central galaxies to halos when a uniform random deviate
is less than the value of Eq. \ref{eq:Ncen} for that halo. We then
compute the number of satellites in each halo hosting a central by
selecting a value from a Poisson distribution with $\lambda=N_{\mathrm{sat}}$
as given by Eq. \ref{eq:Nsat}.

We assign central galaxies to the halo center position and place satellites
at the position of a randomly-selected dark matter particle within
the halo. This eliminates the common assumption of spherical NFW profiles
\citep{1996ApJ...462..563N} and preserves the non-spherical halo
shapes. In a study of how halo occupation assumptions affect galaxy
clustering statistics, \citet{2008ApJ...686...41Z} and \citet{2012arXiv1203.5335V}
both found that assuming spherical halos could decrease the measured
correlation function by up to $\sim10-20\%$ on scales below $r\lesssim1h^{-1}\mathrm{Mpc}$,
compared to the clustering measured with the true halo shapes. They
also reported a reduction in the correlation function on scales of
a few $h^{-1}\mathrm{Mpc}$ if the halos are not correctly aligned
within the overall large scale structure. By placing galaxies at the
location of dark matter particles within each halo, we preserve both
the halo shape and the overall halo alignment. An even better choice
would be to place the galaxies at the locations of subhalos, but our
simulations do not have the resolution to track individual subhalos.
In either case, this method more correctly reproduces the 1-halo/2-halo
transition region of the correlation function \citep{2002PhR...372....1C,2004ApJ...608...16Z}
than distributing galaxies via spherical NFW profiles.

See Appendix \ref{sec:Appendix-HOD parameter systematics} for details
on the physical motivations of these HOD parameters and examples of
how they each affect both the shape of the HOD and the resulting correlation
function.

\subsection{Error estimates}

\label{sub:error estimates}

We perform our fitting procedure below on the projected measurements
(e.g., Eq. \ref{eq:wp_from_xi(rperp,rpll)} and Figure \ref{fig:projected-correlation}),
without incorporating information about the galaxy peculiar velocity
field. To estimate measurement errors, we must shift our mock catalogs
from real into redshift space, place them on the sky, and mask them
with the geometry. In addition, we effectively double the number of
available mock galaxy catalogs by flipping each simulation around
one axis when generating the mock galaxy catalog below, to double
our effective number of simulation boxes from 20 to 40. The flip is
chosen to minimize (NGC) or eliminate (SGC) overlap between the flipped
and non-flipped versions of the box. The overlap in the NGC between
the flipped and non-flipped boxes is less than 10\%, so there should
be minimal signal in the covariance matrix due to this procedure.
Since the geometry of our simulations is a cube, while the survey
geometry is a much more complicated region defined by sectors on the
sky (described at the end of Section \ref{sec:The-Sample}, and represented
graphically in Figure \ref{fig:on-sky-distribution}) and a radial
selection function (i.e., Figure \ref{fig:The-redshift-distribution}),
we remap the periodic simulation cube into a rectangular parallelepiped
via the method and code of \citet{2010ApJS..190..311C}, and then
restrict it to the on-sky mask (see Figure \ref{fig:on-sky-distribution}).

The remapping procedure applies a shear transformation to the periodic
simulation cube. Because the simulation box is periodic, we embed
a new sheared box into the infinite tiling of periodic boxes and take
the new catalog be the points at their new, sheared, positions. This
transformation preserves the simulation volume and number density
and contains each point once and only once (see Figure 2 in \citet{2010ApJS..190..311C}
for a 2d graphical representation of this procedure). This remapping
procedure requires that the box size of the simulations be large enough
so that it preserves large scale structure--an excessively thin sheared
box will remap too many distant points to be in close neighborhoods.
Note that we also must rotate the galaxy velocities (described below)
by a rotation matrix defined by the normalized basis vectors defining
the new sheared box. This places the galaxy velocity vectors, originally
in the coordinate system of the simulation cube, in the coordinate
system of the remapped box.

We then embed the survey geometry into the remapped simulation box
on the sky. We shift the origin of the coordinates of the simulation
box so that its minimum $(x,y,z)$ coordinate corresponds to the minimum
$(x,y,z)$ coordinate for the data (in the NGC, and SGC separately),
effectively placing the mock galaxy box on the sky in the same location
as the galaxy data, but covering a larger area.

To convert our mock catalogs into redshift space, we must assign a
peculiar velocity to each galaxy. Central galaxies are assigned the
bulk velocity of their host halos, which should produce a flattening
of the correlation function along the line of sight due to large scale
motion \citep{1987MNRAS.227....1K}. For satellites, we assign the
peculiar velocity of each galaxy's associated dark matter particle,
as this includes both the bulk halo motion and the infall velocity.
This should result in Fingers-of-God. As our simulation velocities
are ``distance offsets'' relative to the position coordinates, we
can convert the real space coordinates, $r$, into redshift space
coordinates, $s$, via

\begin{equation}
s=r+\hat{r}\cdot\overrightarrow{v},\label{eq:real-to-redshift-with-velocity}
\end{equation}
where $\overrightarrow{v}$ is the galaxy velocity as given above
and $\hat{{r}}$ is the line-of-sight unit vector from the observer's
coordinates $(0,0,0)$ to the galaxy. We then restrict the on-sky
coordinates of the galaxies to the area described in Section \ref{sec:The-Sample}. 

To measure the correlation function on these redshift space mocks,
we use exactly the same binning as was used for the data, measuring
either $\xi(s)$ directly, or $w_{p}(r_{\perp})$ by integrating Eq.
\ref{eq:wp_from_xi(rperp,rpll)}. The error bars shown in Figures
\ref{fig:projected-correlation} and \ref{fig:redshift-space-correlation}
and the covariance matrix used in the MCMC procedure and listed in
Table \ref{tab:correlationfunction} are computed from the standard
deviation of the 20 mock galaxy correlation functions.

\subsection{MCMC fitting}

\label{sub:MockCatalogsandMCMC}

To determine the best-fit HOD parameters, we use a Markov Chain Monte
Carlo (MCMC) method\textbf{ }to fit $w_{p}(r_{\perp})$ and the overall
galaxy number density, as measured from the BOSS data. In particular,
we use the Metropolis-Hastings MCMC formalism \citep{1953JChPh..21.1087M,HASTINGS01041970},
with errors on $w_{p}(r_{\perp})$ from the mock catalogs and assuming
a fixed $15\%$ error on the galaxy density (see below for justification).
This method starts with an initial seed set of values for the HOD
parameters and then, 
\begin{enumerate}
\item Computes a step in a direction determined from the covariance of the
parameters (as discussed in Appendix \ref{sec:Appendix-HOD parameter systematics}).
\item Populates the dark matter simulation with galaxies according to that
HOD.
\item Computes $\xi(r)$ on the resulting mock galaxy catalog.
\item Integrates $\xi(r)$ to obtain $w_{p}(r_{\perp})$ via Eq. \ref{eq:wp_from_xi(r)}.
\item Computes the $\chi^{2}$ between the model and the data.
\end{enumerate}
We accept the new HOD parameters with probability: 
\begin{equation}
\mbox{P}=\min(1,e^{-(\chi_{new}^{2}-\chi_{old}^{2})/2}).\label{eq:accept-value}
\end{equation}
We then iterate this procedure for at least $25,000$ steps, thus
filling out the most likely region of parameter space, which allows
us to easily determine the most likely value and scatter of any statistic
derived from the HOD parameters. For the first run of the MCMC procedure,
we bootstrap with fixed 20\% errors on all correlation function points,
and then use the error estimated from this fit in the following run.
In each subsequent MCMC run, we use the errors on $w_{p}(r_{\perp})$
estimated from the previous run (see Section \ref{sub:error estimates}
for how the error estimates are generated), and iterate this process
until convergence, defined as the mean correlation function of two
runs being the same within the $1\sigma$ errors. The fitted parameters
converge after 3-4 MCMC runs, and the best-fit and mean values, and
estimated errors given in this paper, are taken from the final run.

Although the galaxy number density (Figure \ref{fig:The-redshift-distribution})
varies by $\gtrsim30\%$ across our redshift range, the mock-to-mock
variance in galaxy number density is $\sim2\%$. The survey is not
volume limited, while the mocks are, by construction. In our MCMC
fits, we chose a $15\%$ error on $\bar{{N}}(z)$ to allow for variation
across the redshift range, while still restricting the MCMC code to
values close to the mean of the survey. Through testing, we found
that a tighter restriction, e.g. 5\%, prevents the chains from converging
on a solution. A more complete mock catalog generation process would
assign luminosities and colors to the galaxies and observe them on
the sky in the same manner that the data was observed, resulting in
a mock catalog with a redshift distribution closer to that of the
data \citep[e.g. the method of][]{2009MNRAS.392.1080S}.

Because of variance between simulations, we perform our MCMC fitting
on the ``mean'' simulation, defined as follows. We compute the correlation
function, $\langle\xi(r)\rangle$, of halos with $M_{halo}>10^{13}h^{-1}M_{\odot}$,
and select the simulation with the smallest sum-of-squares difference
from the mean of the 20 correlation functions. This ensures that our
correlation function fitting procedure is not biased high or low due
to a particular simulation box having particularly high or low inherent
halo clustering. Restricting to high mass halos is a rough proxy for
the halos occupied by our galaxies, without any of the randomness
involved in assigning centrals and satellites to halos. An alternative
to choosing the ``mean'' simulation would be to perform the fitting
procedure on all 20 simulations at once, but this is computationally
prohibitive.

We fit the correlation function in the 12 radial bins given in Table
\ref{tab:correlationfunction}, plus the average galaxy number densities
given in Table \ref{tab:galaxy-statistics}. When combined with our
5 parameter model, there are 7 degrees of freedom for the $\chi^{2}$-test
portion of the MCMC. Our mean HOD model has a $\chi^{2}$ of $7.35$
for the NGC and $5.73$ for the SGC.

\begin{table*}
\caption{\label{tab:HOD-parameters}The mean and standard deviation of the
HOD parameters (see Eq. \ref{eq:Ncen} and \ref{eq:Nsat}).}

\centering{}%
\begin{tabular}{cccccc}
\hline 
\hline parameter & mean Full & mean NGC & best-fit NGC & mean SGC & best-fit SGC\tabularnewline
\hline 
$\log_{10}M_{cut}/M_{\odot}$ & $13.25\pm0.26$ & $13.17\pm0.14$ & $13.16$ & $13.09\pm0.09$ & $13.11$\tabularnewline
$\log_{10}M_{1}M_{\odot}$ & $14.18\pm0.39$ & $14.06\pm0.07$ & $14.11$ & $14.05\pm0.09$ & $14.07$\tabularnewline
$\sigma$ & $0.70\pm0.40$ & $0.65\pm0.27$ & $0.741$ & $0.53\pm0.28$ & $0.692$\tabularnewline
$\kappa$ & $1.04\pm0.71$ & $1.46\pm0.44$ & $0.921$ & $1.74\pm0.74$ & $1.26$\tabularnewline
$\alpha$ & $0.94\pm0.49$ & $1.18\pm0.18$ & $1.38$ & $1.31\pm0.19$ & $1.31$\tabularnewline
\hline 
\end{tabular}
\end{table*}

\begin{figure}
\begin{centering}
\includegraphics[width=1\columnwidth]{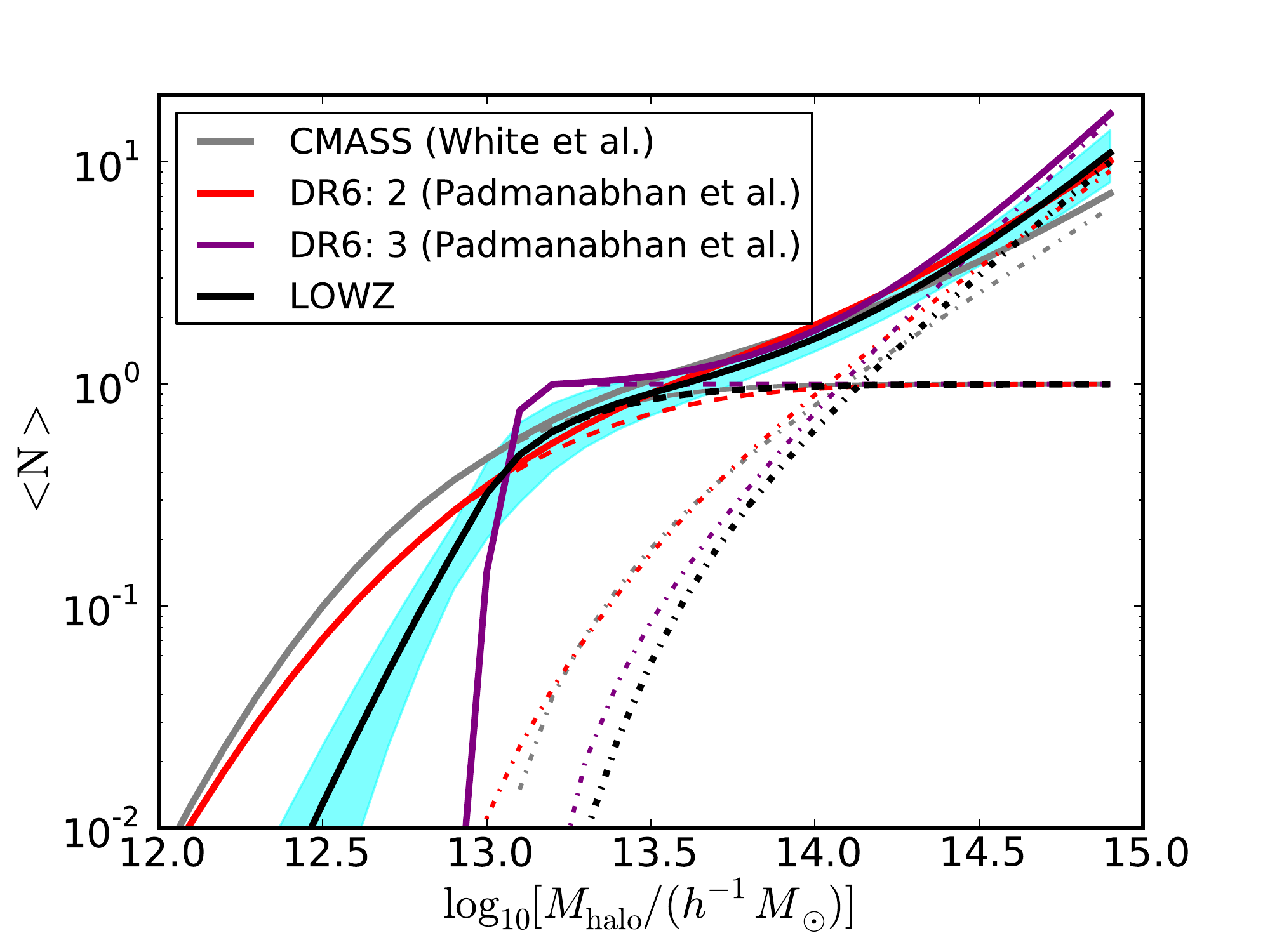}
\par\end{centering}

\caption{\label{fig:Halo-occupation-distribution}Halo Occupation Distributions
(HODs) for LOWZ (NGC+SGC) and CMASS samples compared with HODs for
two samples of LRGs with photometric redshifts from SDSS I/II (``DR6:
2'' with a mean redshift of $0.326$ and ``DR6: 3'' with a mean
redshift of $0.376$). The dashed lines shows the expected number
of centrals, dashed-dotted lines show expected number of satellites,
and solid lines show the total number of galaxies as a function of
halo mass. The shaded region shows the $\pm1\sigma$ variation determined
by our MCMC fitting procedure.}
\end{figure}

Our MCMC procedure produces the mean (calculated from all MCMC steps)
and best-fit (lowest single $\chi^{2}$ during the MCMC procedure)
parameters shown in Table \ref{tab:HOD-parameters}. The parameter
$\kappa$ is poorly constrained, and although $\sigma$ is also not
strongly constrained, it is below the value for the CMASS sample ($0.98\pm0.24$).
We estimate our errors from the mean HOD parameters, not the best-fit,
as the single best-fit value is partly determined by random variations
in the mock correlation functions. Figure \ref{fig:Halo-occupation-distribution}
presents the mean number of galaxies per halo. We also show, for reference,
the HOD determined for the CMASS sample of\textbf{ }\citet{2011ApJ...728..126W},
and two HODs selected from \citet{2009MNRAS.397.1862P} that represent
galaxies with photometric redshifts in a similar redshift and mass
range to those in our current sample. The LOWZ sample clearly lies
between the \citet{2009MNRAS.397.1862P} samples 2 and 3, with a steeper
cutoff for central galaxies than sample 2. The behavior of the HOD
at low halo masses is driven primarily by the amplitude of the correlation
function; the relatively large measured clustering amplitude (compared
with, e.g. CMASS) agrees well with our steep cutoff in the average
number of galaxies in low mass halos. Within the HOD framework, galaxies
with a high bias must occupy high mass halos, resulting in a sharper
turnoff in the central galaxy fraction. For a more detailed comparison
with CMASS, see Section \ref{sec:CMASS comparison}.

\begin{figure}
\begin{centering}
\includegraphics[width=1\columnwidth]{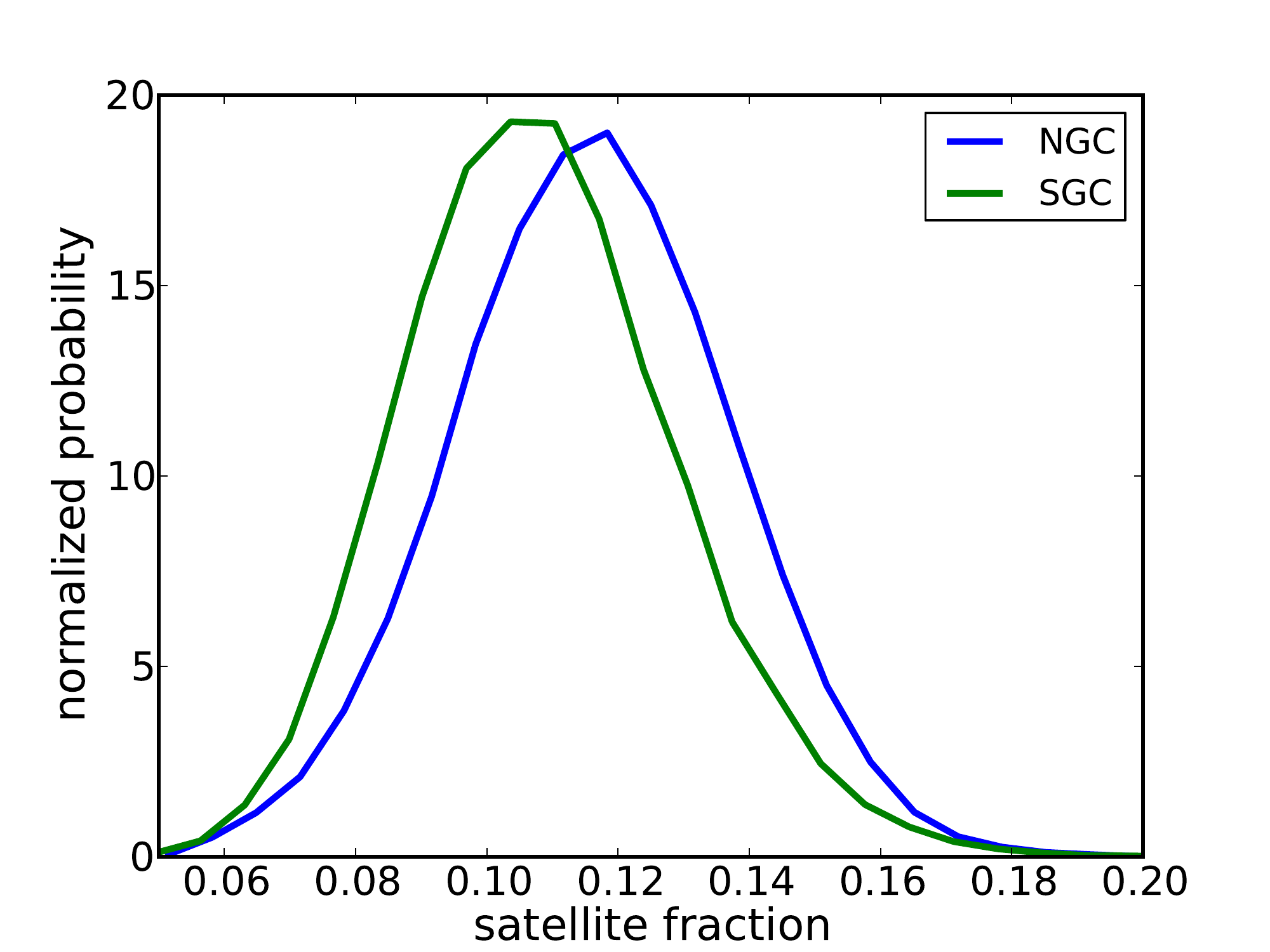}
\par\end{centering}

\caption{\label{fig:satellite-fraction-probability}Probability distribution
(normalized to have an integral of 1) of satellite fractions for NGC
and SGC. }
\end{figure}

Our results suggest that $12\pm2\%$ of NGC and $11\pm2\%$ of SGC
galaxies are satellite galaxies in their halos instead of centrals.
This is comparable with the $\sim10\%$ satellite fraction measured
by \citet{2011ApJ...728..126W} for the higher redshift CMASS sample.
We show the probability distribution function (PDF) of satellite fraction
in Figure \ref{fig:satellite-fraction-probability}. There is considerable
overlap between the NGC and SGC satellite fraction PDFs, and we find
these results to be statistically indistinguishable.

\section{Redshift Space}

\label{sec:Redshift measurement}

\begin{figure*}
\begin{centering}
\includegraphics[width=0.5\linewidth]{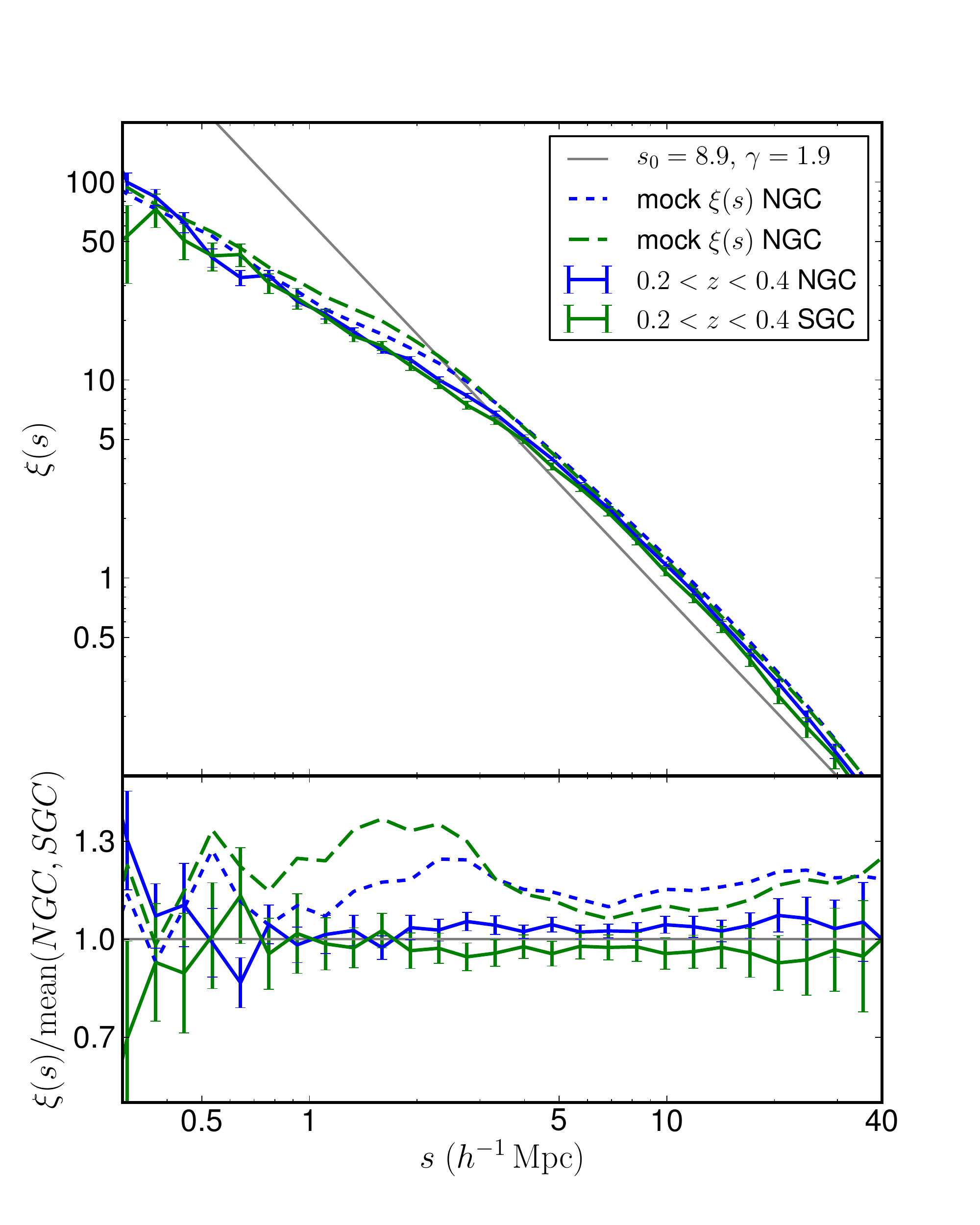}\includegraphics[width=0.5\linewidth]{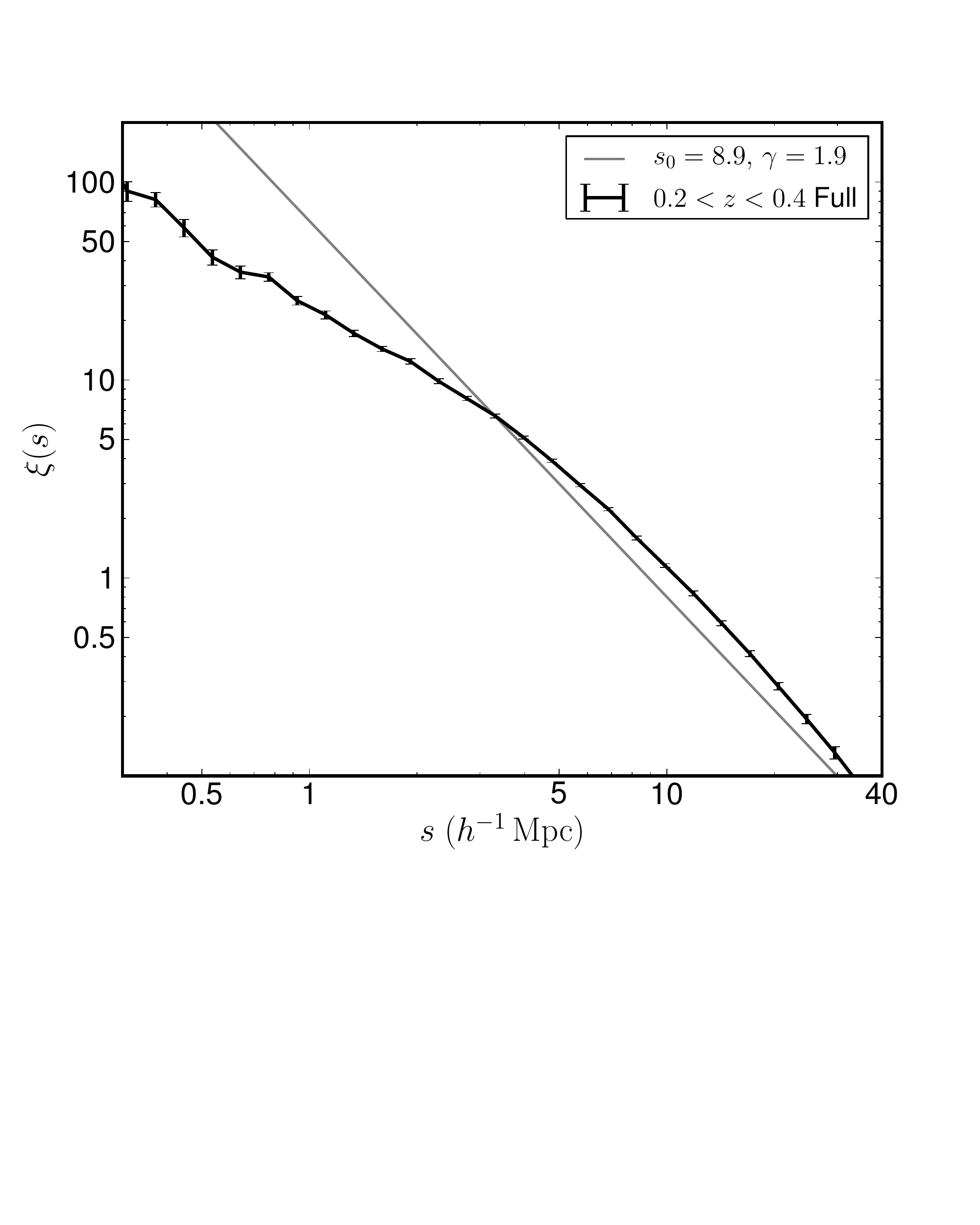}
\par\end{centering}

\caption{\label{fig:redshift-space-correlation}Left: NGC and SGC, data and
mocks. Right: NGC+SGC Full sample.\textbf{ }Upper panels: the redshift
space correlation function for BOSS LOWZ galaxies with redshifts in
the range $0.2<z<0.4$. The thin line shows the same best-fit power
law from Figure \ref{fig:projected-correlation} with $s_{0}=8.9$,
$\gamma=1.9$. The dashed lines show the mean of the 20 mock catalogs
generated from our simulations, using the mean HOD parameters given
in Table \ref{tab:HOD-parameters}. Lower panels: the same curves
as above, but divided by the mean of the NGC and SGC to emphasize
the differences between the various correlation functions. The thin
solid line at $\xi(s)/mean(NGC,SGC)=1$ is not the powerlaw from the
upper panel, but is just shown for reference.}
\end{figure*}

As a test of our HOD fits, we compute the redshift space correlation
function for the data, and for our mean HOD parameters. As our HOD
fitting procedure does not incorporate any information about the velocity
distribution of the galaxy sample, this approach provides a convenient
check of our results. 

Figure \ref{fig:redshift-space-correlation} shows the redshift-space
correlation function, $\xi(s)$. We also plot the same power law from
Figure \ref{fig:projected-correlation} and the mean redshift space
mock catalogs for comparison. The effects of redshift space distortions
are clear here, with a decrease in the correlation function amplitude
at small scales, and an increase at larger scales.

The mocks in this case were not fit to the data, but rather were computed
using the HOD parameters that were fit to the $w_{p}(r_{\perp})$
measurements. The differences between the data $\xi(s)$ and mock
\textbf{$\xi(s)$ }could be due to our requirement that every halo
with satellites must have a central: the lower luminosity galaxies
in our sample may be satellites in halos that do not have a LOWZ galaxy
as their central. The difference may also reflect deviations from
our assumption that the galaxies strictly follow the motion of individual
dark matter particles, as opposed to the subhalos that they truly
do occupy. Further work could expand on this issue by, for example,
making different assumptions about satellites and centrals, selecting
subsets of LOWZ galaxies with different colors and luminosities, or
having the galaxies follow proper subhalos instead of individual dark
matter particles \citep[see, e.g.,][]{2001MNRAS.325.1359S,2001MNRAS.321....1W,2004PhRvD..70h3007S,2006MNRAS.369...68S,2006MNRAS.368...85T}.

\textbf{}

\section{Comparison with previous work}

\label{sec:CMASS comparison}

\textbf{}

\begin{figure}
\begin{centering}
\includegraphics[width=1\columnwidth]{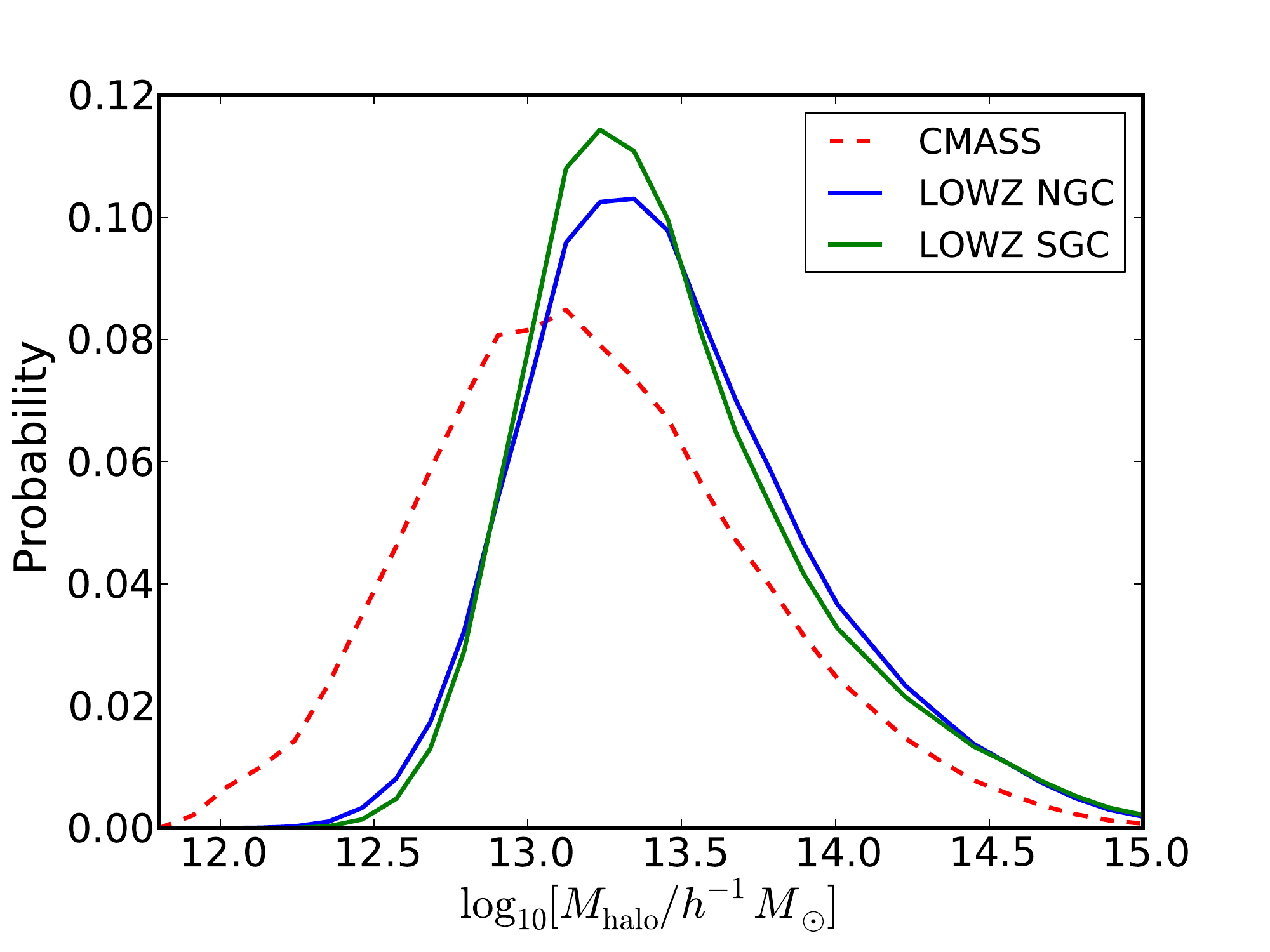}
\par\end{centering}

\caption{\label{fig:Halo-probability}The probability that a galaxy lies in
a halo of a given mass, comparing CMASS and the NGC and SGC LOWZ sample.
Because the HOD for LOWZ galaxies is more sharply truncated than the
CMASS HOD, LOWZ galaxies do not probe halos with masses as small as
the those of the CMASS sample.}
\end{figure}

Figure \ref{fig:Halo-probability} shows a different view on the halo
occupation of these galaxies: the probability that a galaxy lies in
a halo of mass $M$. We compute the mean halo mass to be $5.2\times10^{13}h^{-1}M_{\odot}$
for the LOWZ sample. This Figure clearly shows the sharper halo occupation
cutoff at low halo masses, compared with the CMASS sample, which had
a mean halo mass roughly half as large. This difference is likely
due to the fact that our galaxies are redder than CMASS (Figure \ref{fig:absmags}),
and thus would tend to occupy higher mass halos \citep[see, e.g.,][]{2009MNRAS.399..966S}.
In addition, a galaxy population undergoing dynamical passive evolution
would occupy higher mass halos with cosmic time, as their original
host halos merge to form more massive halos. This is consistent with
the fact that there is little difference between the LOWZ and CMASS
stellar masses (Figure \ref{fig:absmags}): these galaxies have undergone
considerable halo growth, but little to no stellar mass growth since
$z\sim0.5$ \citep[see also][]{2007ApJ...655L..69W,2008ApJ...682..937B,2008MNRAS.387.1045W}.

\begin{figure}
\begin{centering}
\includegraphics[width=1\columnwidth]{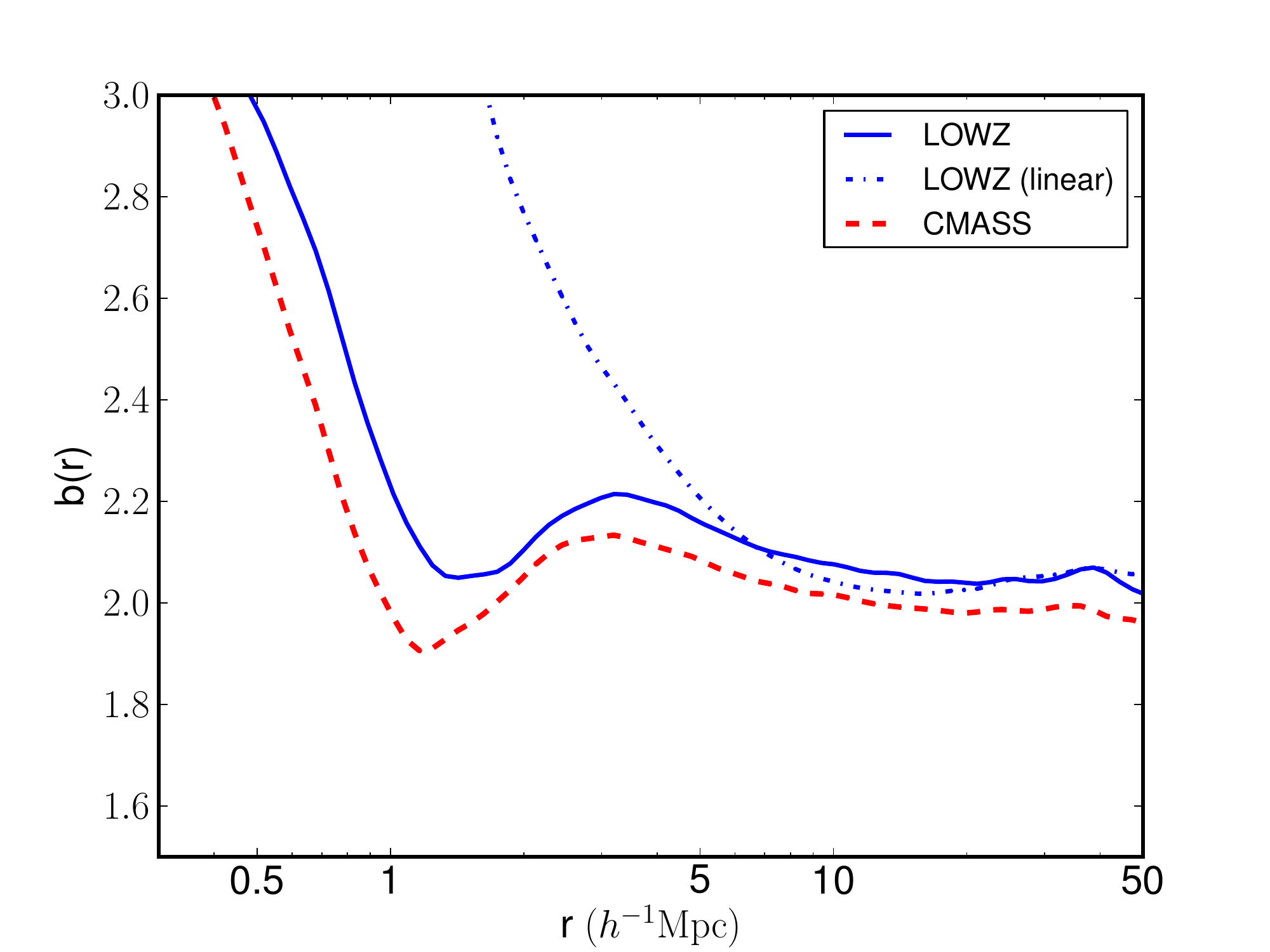}
\par\end{centering}

\caption{\label{fig:bias-scale}The scale dependence of the galaxy bias, $b=\sqrt{\xi_{gal}/\xi_{DM}}$,
for the LOWZ sample. The large-scale bias asymptotes to $\sim2.0$.
The strong increase toward scales below $1h^{-1}\mathrm{Mpc}$ appears
because of the strong clustering of galaxies within halos, while the
bump at the few $h^{-1}\mathrm{Mpc}$ scale is due to one-halo/two-halo
transition. The dashed red line shows the galaxy bias of the CMASS
sample of \citet{2011ApJ...728..126W}, while the dot-dashed blue
line shows the LOWZ galaxy bias relative to the linear theory $\xi_{\mathrm{DM}}$
computed with CAMB.}
\end{figure}

Based on our halo fitting, we can estimate the bias of this galaxy
population with respect to the underlying dark matter distribution,
\[
b(r)=\sqrt{\frac{\xi_{gal}(r)}{\xi_{DM}(r)}},
\]
where $\xi_{gal}(r)$ is the correlation function of the mean galaxy
mock (see Table \ref{tab:HOD-parameters}), and $\xi_{DM}(r)$ is
the correlation function of dark matter in our simulations. We find
a large scale galaxy bias of $\sim2.0$ and show the scale dependence
of the bias in Figure \ref{fig:bias-scale}. The dot-dashed blue line
shows the bias relative to the linear theory correlation function
from CAMB \citep{Lewis:1999bs,Howlett:2012mh}, using the same cosmological
parameters as our simulations. The linear bias differs from the non-linear
bias on small scales but is similar on large scales, as expected.
We find a similar bias to that reported for the CMASS sample of \citet{2011ApJ...728..126W}
(dashed red line).

The measured bias of the full CMASS and LOWZ samples is inconsistent
with pure dynamical passive evolution of CMASS into LOWZ. In a dynamical
passive evolution model in our cosmology, the bias of a galaxy population
evolves like
\begin{equation}
b(z_{0}\rightarrow z)=(b_{z_{0}}-1)\frac{D(z_{0})}{D(z)}+1\label{eq:biasevolution}
\end{equation}
\citep{1996ApJ...461L..65F}. Thus, a bias of $2$ at $z=0.55$ should
evolve to $1.88$ at $z=0.3$. Equivalently, the $z=0.55$ progenitors
of a galaxy population with $b(z=0.3)=2$ should have $b(z=0.55)=2.13$.
\citet{2012arXiv1202.6241T} suggest that the progenitors of SDSS
II LRGs are preferentially located in the redder parts of the color/luminosity
selection space. Together, this suggests that the LOWZ sample could
have passively evolved from a redder, slightly more biased subset
of the CMASS sample. Further work fitting HODs at high redshift and
evolving them to low redshift \citep[e.g.][]{2007ApJ...667..760Z,2007ApJ...655L..69W,2008MNRAS.387.1045W},
as well as selecting subsets of the CMASS sample that may be more
representative of the LOWZ progenitors could clarify this issue.

\citet{2005ApJ...630....1Z} and \citet{2007ApJ...667..760Z} studied
$200,000$ galaxies from the SDSS I/II main sample in the redshift
range $0.01<z<0.23$, with the latter using the same correlation function
measurements but a more complicated form for the HOD. Their samples
were all less luminous than the LOWZ sample, with the most luminous
sample having $-23<M_{r}<-22$ and $0.1<z<0.23$. They found a large-scale
bias factor of $1.91$ for the most luminous sample, with the other
samples having smaller bias. They also measured the mean halo mass
to be higher for fainter red galaxies $\sim2\times10^{14}h^{-1}M_{\odot}$
than for those of intermediate luminosities $\sim1\times10^{14}h^{-1}M_{\odot}$.
Both of these mean halo masses are significantly higher than the mean
halo mass we find for LOWZ galaxies. \citet{2007ApJ...667..760Z}
also added clustering measurements from $30,000$ DEEP2 galaxies in
the redshift range $0.7<z<1.45$. The bias of this sample ranged from
$1.22$ to $1.45$ for the lower ($M_{B}<-19.0$) $ $and higher luminosity
($M_{B}<-20.5$) samples, respectively. They found satellite fractions
for the luminous SDSS and DEEP2 samples (both $\sim10\%$) similar
to our LOWZ results. The results of \citet{2007ApJ...667..760Z} were
updated in \citet{2011ApJ...736...59Z}, incorporating the completed
SDSS I/II main sample galaxy catalog. Their two most luminous samples,
with $M_{r}^{\mathrm{max}}>-22.0$ ($z<0.245$, $\bar{N}=0.5\times10^{-4}h^{3}\mathrm{Mpc}^{-3}$)
and $M_{r}^{\mathrm{max}}>-21.5$ ($z<0.199$, $\bar{N}=2.8\times10^{-4}h^{3}\mathrm{Mpc}^{-3}$),
were the most similar to LOWZ, having $b=2.16\pm0.05$, $f_{\mbox{sat}}=4\%\pm1\%$
and $b=1.67\pm0.03$, $f_{\mathrm{sat}}=9\%\pm1\%$, respectively.

\citet{2009ApJ...707..554Z} used the correlation function measurements
of \citet{2005ApJ...621...22Z} to explore the host halos of $35,000$
luminous red galaxies from the SDSS, with two samples covering $0.16<z<0.36$
and $-23.2<M_{g,z=0.3}<-21.2$, and $0.16<z<0.44$ and $-23.2<M_{g,z=0.3}<-21.8$.
Their $M_{r,z=0.3}<-21.8$ sample has a higher mean halo mass ($\sim10^{14}h^{-1}M_{\odot}$)
compared to LOWZ, while their $M_{r,z=0.3}<-21.2$ has a very similar
mean halo mass ($\sim4.5\times10^{13}h^{-1}M_{\odot}$). Both of these
samples have lower satellite fractions ($\sim6\%$ and $\sim3\%$,
respectively) than our LOWZ sample. A major difference between these
samples and the LOWZ sample is their $3$ times lower number density
of $\sim10^{-4}h^{3}\mathrm{Mpc}^{-3}$, which would impact both the
satellite fraction and halo masses.

\citet{2008MNRAS.387.1045W} fit HOD parameters to color and luminosity
matched SDSS and 2SLAQ \citep{2006MNRAS.372..425C} LRGs. They created
four matched samples by selecting SDSS galaxies with the 2SLAQ color
and magnitude cuts, and 2SLAQ galaxies with the SDSS color and magnitude
cuts, resulting in two samples at low redshift ($z\sim0.2$) and two
at high redshift ($z\sim0.55$). They found that the low redshift
samples had higher mean halo masses than LOWZ ($9.52\times10^{13}h^{-1}M_{\odot}$
and $7.62\times10^{13}h^{-1}M_{\odot}$), but similar satellite fractions
of around $10\%$. On the other hand, their high redshift samples
had lower satellite fractions ($4.7\%$ and $6.2\%$), but more similar
mean halo masses ($6.24\times10^{13}h^{-1}M_{\odot}$ and $4.76\times10^{13}h^{-1}M_{\odot}$).
All of their samples had between $1.5-4$ times lower number density
than the LOWZ sample, ranging from $0.73-1.65\times10^{-4}h^{3}/\mathrm{Mpc}^{-3}$.

\begin{figure}
\begin{centering}
\includegraphics[width=1\columnwidth]{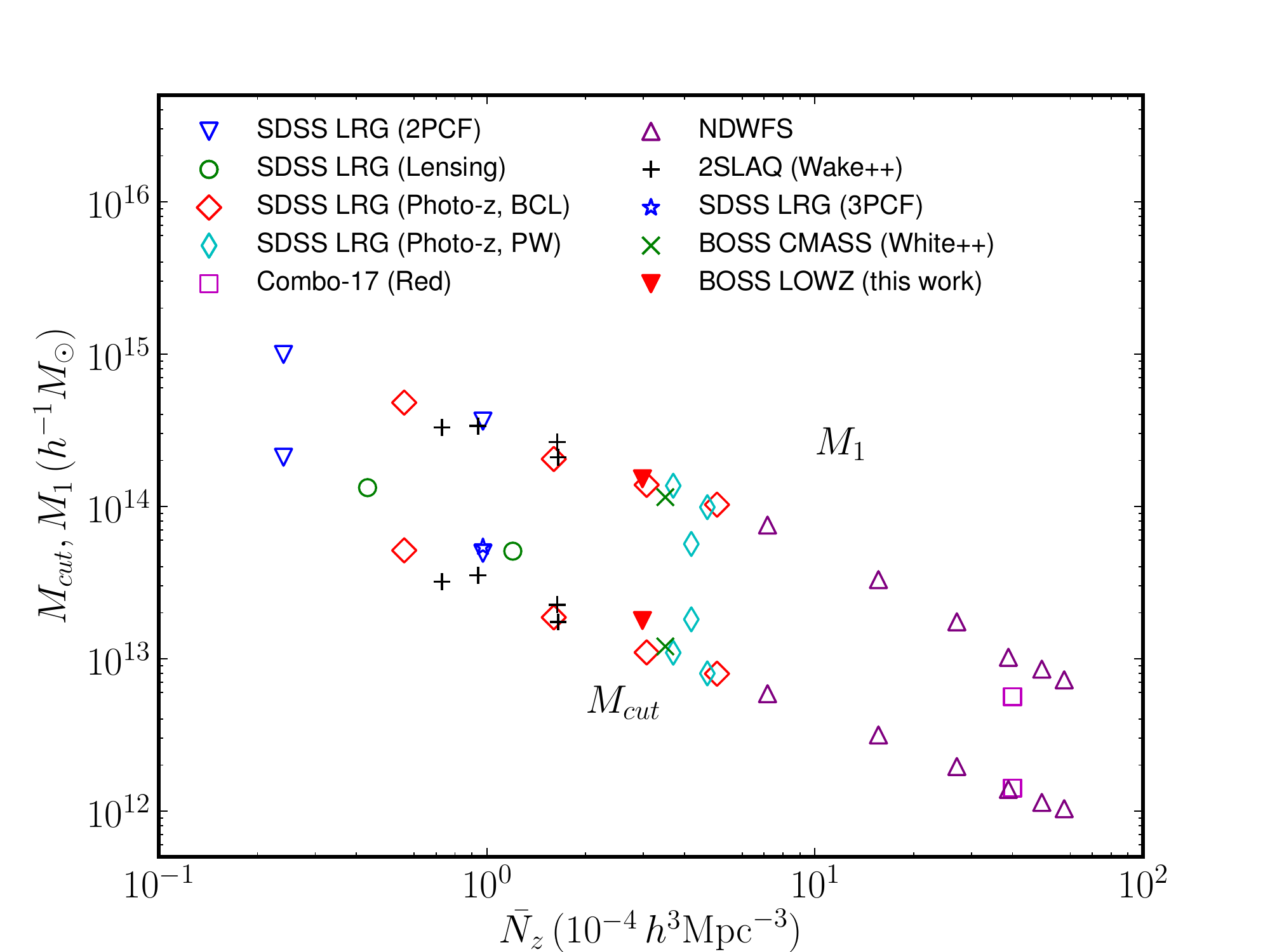}
\par\end{centering}

\caption{\label{fig:Mcut-M1_vs_Nz}$M_{cut}$ and $M_{1}$ vs. $\bar{{N}_{z}}$
for a number of different galaxy correlation function studies. Errorbars,
not shown for clarity, are typically $\sim0.1\,\mathrm{dex}$. The
labels refer to the following studies; SDSS LRG (2PCF): \citet{2009ApJ...707..554Z},
SDSS LRG (Lensing): \citet{2006MNRAS.368..715M}, SDSS LRG (Photo-z,
BCL): \citet{2008MNRAS.385.1257B}, SDSS LRG (Photo-z, PW): \citet{2009MNRAS.397.1862P},
Combo-17 (Red): \citet{2006A&A...457..145P}, NDWFS: \citet{2008ApJ...682..937B},
2SLAQ (Wake++): \citet{2008MNRAS.387.1045W}, SDSS LRG (3PCF): \citet{2007MNRAS.378.1196K},
BOSS CMASS (White++): \citet{2011ApJ...728..126W}. The data in this
plot is provided in the Appendix in Table \ref{tab:Nz_vs_HOD}.}
\end{figure}

Figure \ref{fig:Mcut-M1_vs_Nz} compares our measurements of  HOD
parameters $M_{cut}$ and $M_{1}$ versus galaxy number density, $\bar{{N}_{z}}$,
with other studies from the literature. The HOD parameter estimates
shown in this plot are taken from studies covering a range of redshifts,
sample selection, spectroscopic and photometric catalogs, measurement
techniques, HOD fitting methods, and survey volumes. Our results fit
with the general trend of $M_{cut}$ and $M_{1}$ decreasing for higher
$\bar{{N}_{z}}$. The compilation of different datasets was used in
\citet{2008ApJ...682..937B}, \citet{2011ApJ...728..126W} and \citet{2012arXiv1202.6057N}.
This result is consistent with the observation that galaxy number
density is a rough proxy for the survey's luminosity or stellar mass
limit \citep[e.g.,][]{2005ApJ...631...41T,2005ApJ...633..791Z,2011ApJ...736...59Z}.
For future reference, this data is also compiled in Table \ref{tab:Nz_vs_HOD}.

\section{Conclusions}

\label{sec:Conclusions}

We have described the clustering properties of $\sim80,000$ BOSS
LOWZ galaxies from the SDSS DR9 sample. Our measurements, fitting
procedure, and mock catalogs provide the following properties of this
sample:
\begin{itemize}
\item When working with the LOWZ catalog, we recommend that approximately
the first year's data should not be included in uniform samples because
it was acquired with a shallower selection function. We give a description
of how to restrict the sample to the correct targets. We incorporate
data from SDSS Legacy to provide about one third of our redshifts,
and provide a description of how to include these data when generating
large scale structure catalogs.
\item The clustering of the NGC and SGC samples differ by less than $1\sigma$,
even though the south has a $\sim10\%$ higher number density because
of variations in photometry. We present correlation function and fitted
parameter values for the Full NGC+SGC sample and for both the NGC
and SGC separately.
\item The LOWZ sample has a higher correlation function amplitude ($r_{0}\sim8.9$),
but a similar bias ($b\sim2.0$) compared to the higher redshift CMASS
sample.
\item The best-fit and mean HOD both result in a higher average host halo
mass ($5.2\times10^{13}h^{-1}M_{\odot}$) and steeper mass cutoff
for the LOWZ sample compared to the CMASS sample, but well within
the range of similar photometric galaxy samples from SDSS I/II.
\item Our HOD fits result in a satellite fraction of $\sim11\%$ for the
LOWZ sample. 
\item The LOWZ sample is broadly consistent with being passively evolved
analogs of a subset of the higher redshift CMASS sample.
\end{itemize}

\section{Acknowledgments}

Funding for SDSS-III has been provided by the Alfred P. Sloan Foundation,
the Participating Institutions, the National Science Foundation, and
the U.S. Department of Energy Office of Science. The SDSS-III web
site is http://www.sdss3.org/.

SDSS-III is managed by the Astrophysical Research Consortium for the
Participating Institutions of the SDSS-III Collaboration including
the University of Arizona, the Brazilian Participation Group, Brookhaven
National Laboratory, University of Cambridge, Carnegie Mellon University,
University of Florida, the French Participation Group, the German
Participation Group, Harvard University, the Instituto de Astrofisica
de Canarias, the Michigan State/Notre Dame/JINA Participation Group,
Johns Hopkins University, Lawrence Berkeley National Laboratory, Max
Planck Institute for Astrophysics, Max Planck Institute for Extraterrestrial
Physics, New Mexico State University, New York University, Ohio State
University, Pennsylvania State University, University of Portsmouth,
Princeton University, the Spanish Participation Group, University
of Tokyo, University of Utah, Vanderbilt University, University of
Virginia, University of Washington, and Yale University. 

This work was supported in part by the facilities and staff of the
Yale University Faculty of Arts and Sciences High Performance Computing
Center, the National Energy Research Scientific Computing Center,
the Shared Research Computing Services Pilot of the University of
California, and the Laboratory Research Computing project at Lawrence
Berkeley Laboratory.

\appendix{}

\section{HOD parameters}

\label{sec:Appendix-HOD parameter systematics}

The central galaxy occupation function in Eq. \ref{eq:Ncen} allows
for a log-normal scatter between galaxy luminosity and halo mass with
logarithmic dispersion $\sigma$, so that the occupation of central
galaxies rises smoothly from zero to one. The satellite occupation
function of Eq. \ref{eq:Nsat} is a power-law of slope $\alpha$ smoothly
truncated at low halo masses, normalized by the mass $M_{1}$ at which
halos have approximately one satellite on average. The parameter $\kappa$
allows different cutoffs in the central and satellite occupations,
and it ensures that the normalization of the satellite occupation
is not driven artificially by the behaviour of low-occupation halos.
Occupation functions of this form provide a good fit to the theoretically
predicted occupations of galaxy samples limited by stellar mass or
luminosity \citep{2004ApJ...609...35K,2005ApJ...633..791Z}. Our choice
of HOD represents a step function with a lower mass threshold for
the number of centrals (Eq. \ref{eq:Ncen}), smoothed out and increasing
slowly to a probability of 1 over some range (the complementary error
function). Similarly, as we expect the number of satellites (Eq. \ref{eq:Nsat})
to increase at high halo mass, we assume a power law for the number
of satellites, with a minimum mass cutoff ($\kappa M_{cut}$) to prevent
small halos from having satellites. Because of the color selection
and the modest dependence of luminosity threshold on redshift in the
BOSS LOWZ sample, we may expect the dispersion $\sigma$ to be somewhat
larger than it would be for a sharply thresholded sample. See, e.g.
\citet{2002ApJ...575..587B,2004MNRAS.347..813H,2005ApJ...621...22Z,2005ApJ...633..791Z,2007ApJ...667..760Z,2008MNRAS.387.1045W}
for other choices of HOD parameterization, some of which are analogous
to ours.

As a rough guideline, the parameters affect the resulting HOD, and
thus the correlation function in the following ways. $M_{cut}$ determines
the characteristic minimum mass of halos in which a galaxy is allowed,
and thus affects the overall amplitude of the correlation function,
with smaller values of $M_{cut}$ resulting in a lower amplitude.
The quantity $\sigma$ changes the shape of the ``central galaxy
probability function'', with smaller $\sigma$ resulting in fewer
galaxies populating lower mass halos with lower large-scale bias,
and thus a higher correlation function amplitude on large scales.
$M_{1}$ affects the number of satellites, with smaller $M_{1}$ producing
more satellites, and thus a higher amplitude for the correlation function
overall, and a steeper slope at small radii. Increasing $\alpha$
increases the number of satellites, adding power on small scales,
and affecting the one-halo/two-halo transition. Varying $\kappa$
has a small effect, tuning the mass at which satellites are allowed
and slightly altering the small-scale correlation function. These
parameters have some degeneracy, with $\sigma$ and $M_{cut}$, and
$\alpha$ and $M_{1}$ positively correlated, and $M_{1}$ and $M_{cut}$
negatively correlated; we find little dependence on $\kappa$ within
the range we sample, $0<\kappa<3$. Other studies that explore the
relation of the HOD parameters with the resulting correlation function
include \citet{2002ApJ...575..587B,2009ApJ...707..554Z,2011ApJ...738...22W}.

\begin{figure*}
\begin{centering}
\includegraphics[width=0.8\linewidth]{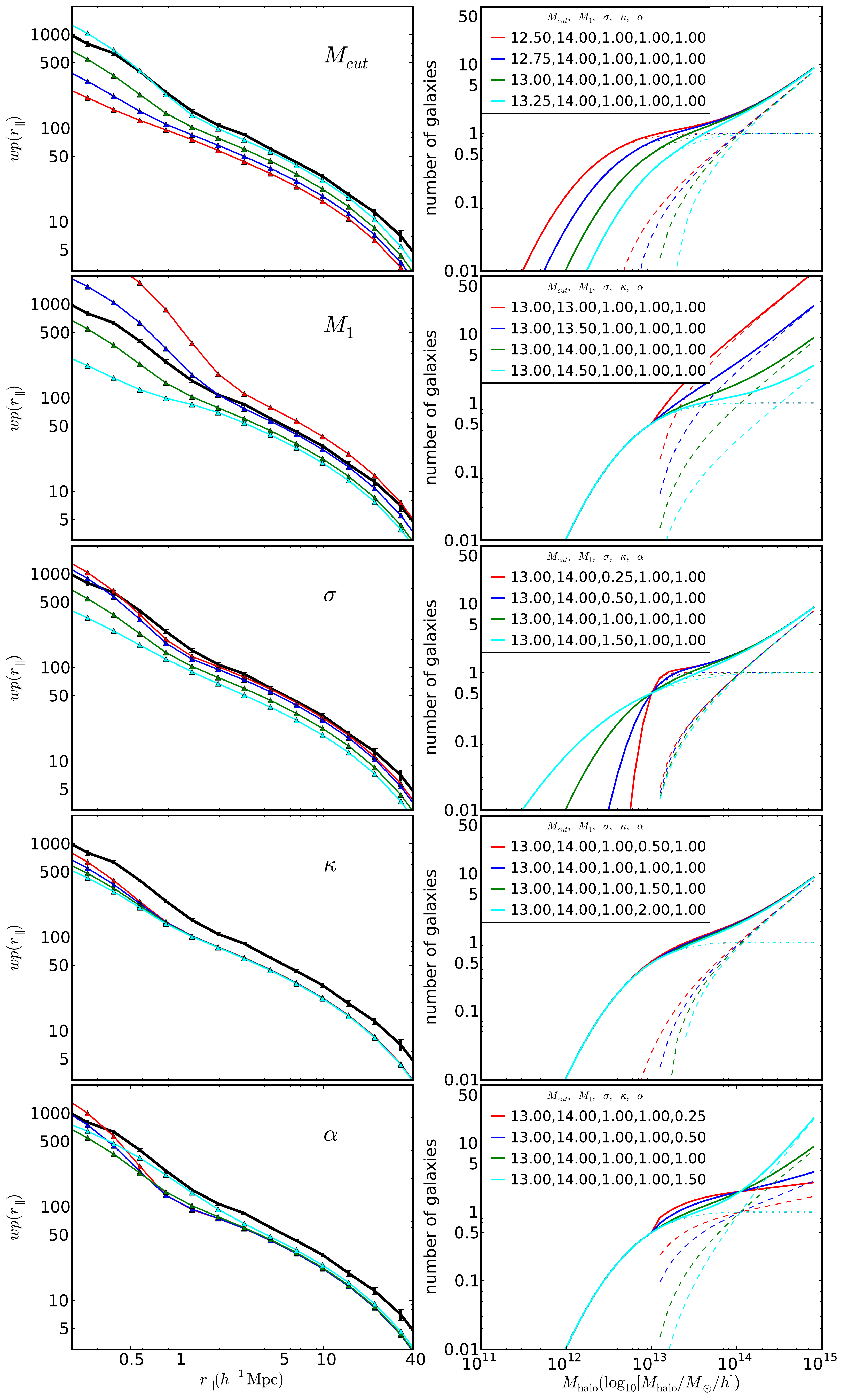}
\par\end{centering}

\caption{\label{fig:HOD-parameter-covariance}Variation of the correlation
function and HOD as individual HOD parameters change. From the top
panel to the bottom panel, the parameter that is varied is $M_{1}$,
$M_{cut}$, $\sigma$, $\kappa$, $\alpha$. The thick black line
is the NGC correlation function described in this paper.}
\end{figure*}

Figure \ref{fig:HOD-parameter-covariance} shows how the correlation
function, $w_{p}(r_{p})$, and the satellite and central halo functions
vary as each individual HOD parameter is changed over a range of reasonable
values. Each panel includes one set of curves with the values $M_{1}=13.00$,
$M_{cut}=14.00$, $\sigma=1.00$, $\kappa=1.00$, $\alpha=1.00$ as
a fiducial. The degeneracies in the correlation functions between
the parameters are visible, as well as the effects that each has on
the shape and amplitude of the resulting correlation function.

The values of $M_{cut}$ and $M_{1}$ from previous studies used in
Figure \ref{fig:Mcut-M1_vs_Nz} are given in Table \ref{tab:Nz_vs_HOD}. 

\begin{table}
\begin{centering}
\caption{\label{tab:Nz_vs_HOD}$M_{cut}$ and $M_{1}$ versus $\bar{N}_{z}$
from various studies}

\par\end{centering}

\begin{centering}
\begin{tabular}{cccc}
\hline 
\hline$\bar{N}_{z}$ & $\log_{10}M_{cut}$ & $\log_{10}(M_{1})$ & Sample\tabularnewline
\hline 
9.730E-5 & 13.6907 & 14.5587 & 0\tabularnewline
2.400E-5 & 14.3217 & 14.9967 & 0\tabularnewline
1.200E-4 & 13.7055 & 0.0000 & 1\tabularnewline
4.330E-5 & 14.1224 & 0.0000 & 1\tabularnewline
9.730E-5 & 13.7328 & 0.0000 & 7\tabularnewline
9.400E-5 & 13.5464 & 14.5278 & 6\tabularnewline
1.640E-4 & 13.3540 & 14.4215 & 6\tabularnewline
5.753E-3 & 12.0150 & 12.8610 & 5\tabularnewline
4.916E-3 & 12.0560 & 12.9310 & 5\tabularnewline
3.888E-3 & 12.1420 & 13.0090 & 5\tabularnewline
2.708E-3 & 12.2920 & 13.2420 & 5\tabularnewline
1.562E-3 & 12.4990 & 13.5190 & 5\tabularnewline
7.171E-4 & 12.7700 & 13.8770 & 5\tabularnewline
5.030E-4 & 12.9008 & 14.0108 & 2\tabularnewline
3.070E-4 & 13.0408 & 14.1408 & 2\tabularnewline
1.600E-4 & 13.2708 & 14.3108 & 2\tabularnewline
5.600E-5 & 13.7108 & 14.6808 & 2\tabularnewline
3.700E-4 & 13.0390 & 14.1350 & 3\tabularnewline
4.700E-4 & 12.9029 & 13.9942 & 3\tabularnewline
4.200E-4 & 13.2578 & 13.7528 & 3\tabularnewline
7.300E-5 & 13.5057 & 14.5182 & 6\tabularnewline
1.650E-4 & 13.2408 & 14.3228 & 6\tabularnewline
4.000E-3 & 12.1500 & 12.7500 & 4\tabularnewline
3.500E-4 & 13.0800 & 14.0600 & 8\tabularnewline
2.981E-4 & 13.2500 & 14.1800 & 9\tabularnewline
\end{tabular}
\par\end{centering}

\centering{}{\footnotesize Sample numbers refer to the following works:
0: SDSS LRG (2PCF), \citet{2009ApJ...707..554Z}; 1: SDSS LRG (Lensing),
\citet{2006MNRAS.368..715M}; 2: SDSS LRG (Photo-z, BCL), \citet{2008MNRAS.385.1257B};
3: SDSS LRG (Photo-z, PW), \citet{2009MNRAS.397.1862P}; 4: Combo-17
(Red), \citet{2006A&A...457..145P}; 5: NDWFS, \citet{2008ApJ...682..937B};
6: 2SLAQ (Wake++), \citet{2008MNRAS.387.1045W}; 7: SDSS LRG (3PCF),
\citet{2007MNRAS.378.1196K}; 8: BOSS CMASS (White++), \citet{2011ApJ...728..126W};
9: this work.}
\end{table}

\bibliographystyle{mn2e}
\bibliography{plbib,bib/clustering_method,bib/hod,bib/sdss,bib/boss,bib/cosmology,bib/mcmc,bib/galaxies,bib/sdss-tech}

\end{document}